\documentclass[11pt]{article}

\usepackage[margin=1.25in]{geometry}

\usepackage{setspace}

\usepackage[labelfont=bf]{caption}

\usepackage[numbers,sort&compress]{natbib}

\usepackage{amsmath,amssymb,bm}
\usepackage{siunitx}
 \usepackage{xcolor}
\usepackage{graphicx}
\usepackage{multicol}
\usepackage{physics}
\renewenvironment{abstract}
	{\quotation}
	{\endquotation}

\date{}

\usepackage{url}

\newcommand{\blue}[1]{\textcolor{blue}{#1}}

\usepackage{upgreek}

\usepackage{authblk} 
\usepackage[utf8]{inputenc}

\title{\bf Scallop Theorem for Swimming in Anisotropic Fluids}

\author[1,2,$\dagger$]{Mojtaba Rajabi}
\author[3,4,$\dagger$]{Enej Caf}
\author[1]{Qi Xing Zhang}
\author[1]{Stephen A. Crane}
\author[1]{Zhijian Zhou}
\author[3,5]{Miha Ravnik}
\author[4]{Gareth P. Alexander}
\author[3,5,$\ast$]{Žiga Kos}
\author[1,$\ast$]{Kathleen J. Stebe}

\affil[1]{Department of Chemical and Biomolecular Engineering, University of Pennsylvania, Philadelphia, PA 19104, USA}
\affil[2]{Advanced Materials and Liquid Crystal Institute, Kent State University, Kent, OH 44242, USA}
\affil[3]{Faculty of Mathematics and Physics, University of Ljubljana, Ljubljana, Slovenia}
\affil[4]{Department of Physics, University of Warwick, Coventry, CV4 7AL, United Kingdom}
\affil[5]{Department of Condensed Matter Physics, Jožef Stefan Institute, Ljubljana, Slovenia}

\affil[$\ast$]{Corresponding authors. Email: \texttt{ziga.kos@fmf.uni-lj.si}, \texttt{kstebe@seas.upenn.edu}}
\affil[$\dagger$]{These authors contributed equally to this work.}

\begin{document}

\maketitle

\begin{abstract}
In isotropic fluids like water, microswimmers have evolved to swim despite their tiny size. As described by Purcell in his Scallop Theorem, reciprocal motions cannot drive swimming absent inertia, as is typical at micrometer length scales. Thus, such swimmers have evolved complex structures that perform non-reciprocal swim strokes to propel. In structured fluids, the broken symmetries arising from fluid anisotropy fundamentally alter the rules of propulsion, enabling the creation of active suspensions even with the simplest isotropic swimmers. We combine experiments on magnetically rotated colloids in nematic liquid crystals with analytic solutions to formulate propulsion strategies for microswimmers. We find that spherical colloids swim effectively under continuous rotation and under reciprocal forcing. Swim strokes that are ineffective in isotropic fluids are highly effective in nematic liquid crystals. In light of these observations, the Scallop Theorem is extended for structured fluids.
\end{abstract}

The field of active colloids has undergone meteoric growth in the past decade, touching fields ranging from biology to physics, with application to micro robotics, reconfigurable materials, and biomimetic systems. Diverse strategies are developed to generate colloid motion, including body displacements or “swimming strokes” that are effective for propulsion. The movement of swimmers at low Reynolds numbers in Newtonian fluids is determined by dominant viscous forces. Neglecting inertia results in instantaneous responses and the corresponding dynamic equations become time-independent. Thus, as famously described by Purcell's Scallop Theorem~\cite{purcell2014life}, in order to propel or swim, the micro-swimmer has to perform a non-reciprocal sequence of movements. This is a strong constraint which limits the variety of swimming apparatuses that are effective, e.g. flagella~\cite{elgeti2015physics} \emph{in vivo} and biomimetic designs \emph{in vitro}~\cite{Najafi2004,avron2005pushmepullyou,Lauga2009}. To lift this constraint one has to exploit the surrounding fluid. It is well established that rate-dependent fluids allow for speed enhancement of swimmers~\cite{spagnolie2023swimming,elfring2014theory} and even enable propulsion of one-hinged scallops~\cite{Qiu2014} and counter-rotating dumbells~\cite{yasuda2020reciprocal,datt2018two,pak2012micropropulsion,kroo2022freely,puente2019viscoelastic,eberhard2023reciprocal}. However, there are no guidelines regarding swimming in structured complex fluids with underlying molecular order.

\begin{figure}[h!] 
	\centering
	\includegraphics[width=0.5\textwidth]{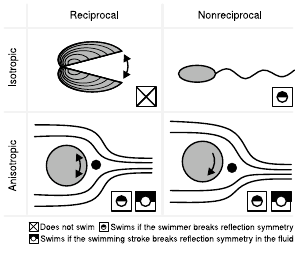} 
	\caption{ \textbf{~Generalization of the Scallop theorem for anisotropic fluids.} 
    In isotropic fluids at low Reynolds numbers, scallops performing reciprocal motions cannot swim. Rather, complex non-reciprocal swimming strokes with broken reflection symmetry are needed to achieve propulsion, typically requiring specially evolved apparatus like a flagellum. We formulate criteria for swimming in nonlinear anisotropic fluids, and demonstrate that even reciprocal swimmers propel, provided either that their swimming strokes break reflection symmetry in the fluid or that they disrupt the symmetry in the surrounding fluid structure. This latter criterion dramatically expands the types of swimmer motions and body shapes that generate thrust. Both requirements are also present for swimmers using non-reciprocal swimming strokes. Thus, effective non-reciprocal swimmers include those that could swim in isotropic fluids, and can be extended to include the simplest of isotropic swimmers, like spherical colloids, whose rotation distorts the fluid’s anisotropic order.} 
	\label{figure 1} 
\end{figure}

Nematic order arises in a variety of complex fluids, ranging from biological systems  such as bacterial colonies~\cite{hartmann2019emergence}, epithelia tissues~\cite{saw2017topological}, and actin fibers during embriogenesis~\cite{maroudas2021topological}, to liquid crystalline fluid domains formed by nucleic acids including RNA~\cite{todisco2018nonenzymatic}  and DNA~\cite{fraccia2015abiotic}. Nematic liquid crystals offer precise experimental control over orientational order and have become the primary model system for investigating colloidal particles in anisotropic environments. Their intrinsic molecular organization has been harnessed to direct the assembly of colloidal particles into crystalline lattices ~\cite{musevic2006two,mundoor2016triclinic}, amorphous solids~\cite{wood2011self}, or even for functional materials~\cite{yao2022nematic,miller2014design}.
At equilibrium, Brownian colloidal particles display anomalous diffusion in structured fluids~\cite{turiv2013effect}. The fluid's ordered structure can be also used to generate distinctively non-equilibrium materials like propelled colloidal particles, for example rotating disks~\cite{yao2022topological} and microrobots~\cite{yao2022nematic}, pulsating bubbles~\cite{kim2024symmetrically}, bacteria-laden droplets~\cite{rajabi2021directional}, microrods with optically driven surface reconfigurations of molecular order~\cite{eremin2015optically,tavera2025quorum} or particles driven by oscillating electric fields~\cite{senyuk2024out} and liquid crystal enabled electrophoresis~\cite{Lavrentovich2010,hernandez2014reconfigurable}. 
Despite such examples of colloidal propulsion in nematic fluids, there are no clear rules regarding the constraints imposed on effective swim strokes by structured non-Newtonian fluids.

In this paper, we establish clear rules for propulsion in structured anisotropic fluids (Fig. \ref{figure 1}), which we demonstrate via analysis of colloidal particles in nematic liquid crystals, one of the simplest realizations of structured fluids, comprising elongated  molecules aligned along a common axis called the director $\vb{n}$. The director field is anchored at the surface of the particles and can provide broken symmetries even for spherically symmetric inclusions. Using rotating ferromagnetic spheres in nematic fluids, we demonstrate that propulsion due to reciprocal and non-reciprocal swimming strokes scales quadratically or linearly with the driving frequency $\mathit\Omega$, respectively, and show the dependence of the swimming velocity on the fluid's viscosity parameters. 

\subsection*{Preparation of microswimmers}
Our experimental setup consists of a planar cell filled with 5CB nematic liquid crystal (Fig.~\ref{figure 2}A). The cell is formed by two bounding plates treated to generate a uniformly oriented planar nematic director field. Upon introduction of a spherical colloidal particle with  perpendicular anchoring, the liquid crystal minimizes its free energy around the colloid subject to an important topological constraint. In this setting, the spherical colloid forms a defect of topological charge $+1$ in the director field; a companion topological defect must emerge spontaneously near the particle in the otherwise uniform director. Two topologically equivalent configurations can form. In one arrangement, the defect is a Saturn ring, a disclination line with winding number $-1/2$ at the colloid's equator~\cite{Gu2000}. In the other arrangement, a dipolar configuration emerges, with a ``hedgehog'' companion defect of topological charge $-1$ (Fig.~\ref{figure 2}A,B) ~\cite{Poulin1997}. The nanometric defects scatter light because of the nematogen's birefringence, and are observable under optical microscopy (Fig.~\ref{figure 2}B).

\subsection*{Non-reciprocal swimming strokes}
\begin{figure} [!htbp]
	\centering
	\includegraphics[width=0.5\textwidth]{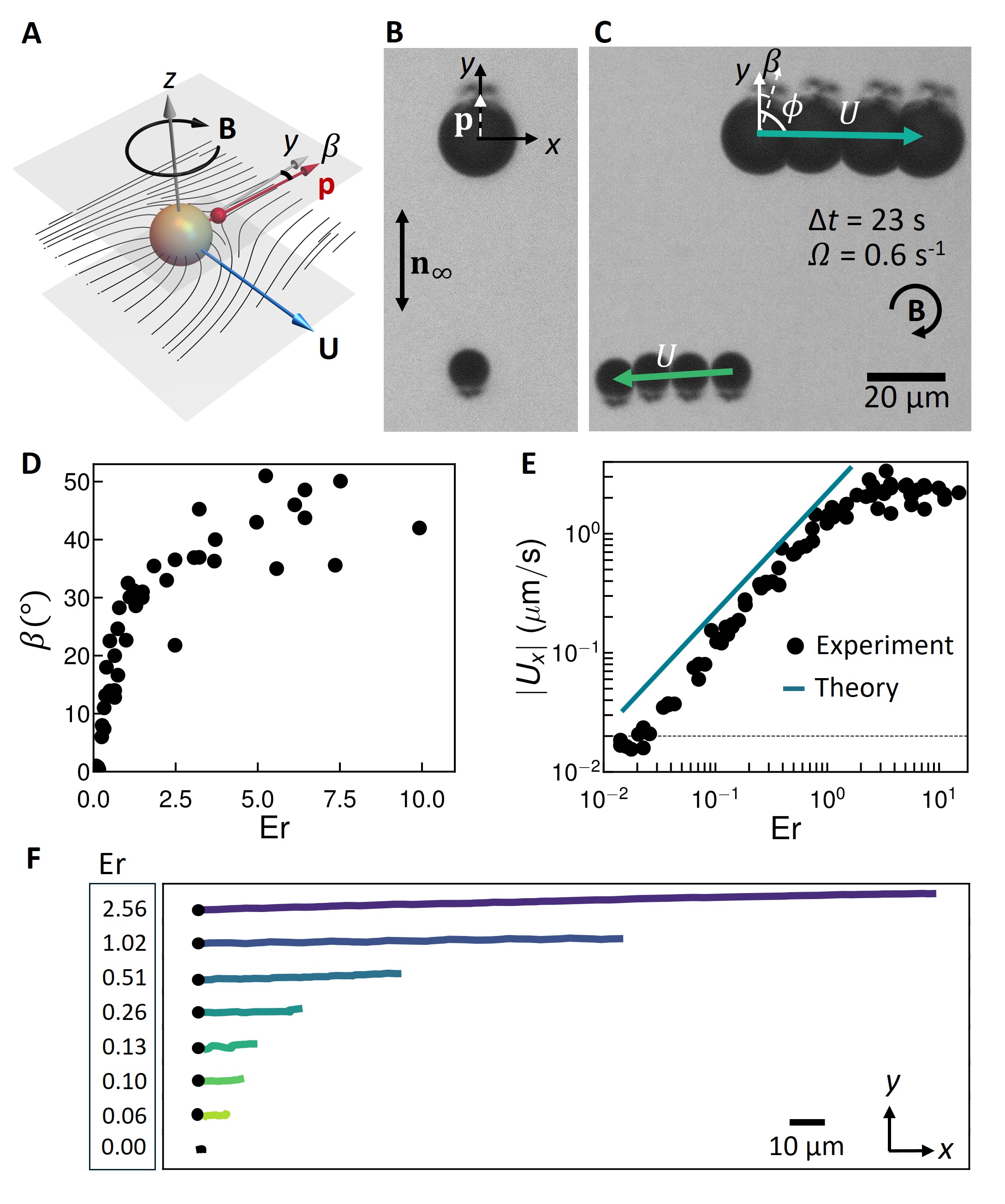} 
	\caption{\textbf{~Swimming of a rotating spherical colloid in a nematic liquid crystal}
		 (\textbf{A}) Schematic of the nematic director field near a sphere accompanied by a topological point defect, confined between two plates with a gap of $50~\mathrm{\mu m}$. (\textbf{B}) Microscopic image of spheres in uniformly aligned 5CB at equilibrium, accompanied by topological point defects. The vector $\mathbf{p}$ connects the center of the sphere to the point defect. (\textbf{C}) Overlay of the spheres and their defects as they swim perpendicular to the far field director driven by rotational forcing. The colored arrows represent the spheres' trajectories. $\Delta t$ is the time interval between two consecutive overlaid images.  
         (\textbf{D}) Angle $\beta$ between $\mathbf{p}$ and the $y$-axis as a function of $\mathrm{Er}$. (\textbf{E}) Swimming speed of rotating spheres as a function of Er. The solid blue line represents the theoretical prediction (Eq.~\ref{eq:propulsion}), which has a slope of 1. The horizontal line represents a limit below which drift dominates; this drift is attributed to hydrodynamic interactions, gradients of the magnetic field, and weak flows present in the cell. All material parameters used to plot the theoretical prediction are reported in the Methods section. (\textbf{F}) Trajectories of a rotating sphere at different Er over time interval $t=30~s$.}
	\label{figure 2} 
\end{figure}

We first focus on the dipolar arrangement, which is the most stable configuration for a micrometer scale particle. The hedgehog companion defect can form with equal probability at either pole of the particle aligned with the far field nematic director $\mathbf {n_\infty}$. We define a Cartesian coordinate at the sphere's center, so that the $y$-axis points in the direction of the hedgehog defect and the $z$-axis points along the normal of the bounding plates of the cell~(Fig.~\ref{figure 2}A). The far field  $\mathbf {n_\infty}$ is aligned with the $y$-axis. At equilibrium, the director field is symmetric in $x$, and highly asymmetric in $y$~(Fig.~\ref{figure 2}A). Steady rotation of the colloid around the $z$-axis generates viscoelastic stresses that depend on the nematogen arrangement and can drive the colloid translation in the $x$-direction~(Fig.~\ref{figure 2}C). Moreover, rotation of the colloid generates disturbances in the director field and broken symmetries along the $x$-axis ~(Fig.~\ref{figure 2}C), changing the magnitude and direction of propulsion.

We rotate ferromagnetic spheres around the $z$-axis using a weak external magnetic field generated by two pairs of Helmholtz coils. As a control, we first rotate these spheres in water; as expected, the external field exerts a torque on the particle whose rotation generates purely axisymmetric viscous stresses in the surrounding fluid. No net forces arise to drive translation, consistent with Purcell's Scallop Theorem for isotropic fluids. Indeed, by tracking the rotated particle's position over time, we find that its displacements in water are consistent with Brownian diffusion. (Fig. \ref{Sup water rotation} and Movie S1).

\indent 
We then introduce the ferromagnetic spheres treated to have perpendicular anchoring into the planar liquid crystal-filled cell and rotate it around the $z$-axis. The disturbance in the nematic director field is readily apparent, as the hedgehog defect is displaced from its equilibrium position by an angle $\beta$ towards the direction of rotation (Fig.~\ref{figure 2}C).
Furthermore, the particle ``swims'' effectively, translating primarily along the $x$-axis (Fig.~\ref{figure 2}C and Fig. \ref{Sup 5CB rotation} and Movie S2), with its direction of motion determined by the sense of rotation; a sphere rotating clockwise (counter-clockwise) moves in the $\mathbf{x}$ ($\mathbf{-x}$) direction (Fig.~\ref{figure 2}C).

The sphere's swimming behavior depends on the Ericksen number $\mathrm{Er} = \mathit\Omega\tau$ where $\mathit\Omega$ is sphere's angular frequency and $\tau = R^2\gamma_1/K$ is the visco-nematic relaxation timescale. In this expression, $K$ is the average bulk elastic coefficient of the nematic and $\gamma_1$ is the rotational viscosity. The spheres in our experiments  are polydisperse, with radii ranging from ~$R\approx5 - 12 $~$\upmu$m. In order to compare spheres with similar confinement, we report $\beta$, trajectories,  and swimming velocities versus $\mathrm{Er}$ for particles with radii ~$R=9 - 12 $~$\upmu$m. As $\mathrm{Er}$ increases, so do the broken symmetries in the director, made evident by increases in the offset angle of the dipolar defect $\beta$ (Fig.~\ref{figure 2}D). The sphere's propulsion speed is linearly proportional to $\mathrm{Er}$~(Fig.~\ref{figure 2}E,F). Deviations from this linear dependence at small $\mathrm{Er}$ can be attributed to Brownian diffusion and weak drift ~(Fig.~\ref{figure 2}E and Fig. \ref{Sup 5CB rotation}). 

To describe the swimming behavior of rotating colloidal spheres, we first calculate the flow-induced distortions of the director field, which affect the Ericksen-Leslie stress tensor. Motivated by~\cite{neville2024controlling}, we formulate the analytical description of the swimming mechanism within an extended Lorentz reciprocal relation designed for a general swimmer in a complex fluid. This framework relates a complex fluid problem (propulsion speed ${\bf U}$, angular velocity ${\bm \Omega}$, torque ${\bf T}$, force ${\bf F}$, and complex fluid stress tensor ${\bm \Sigma}$) with an auxiliary isotropic fluid problem (propulsion speed ${\bf U}^{\prime}$, angular velocity ${\bm \Omega}^{\prime}$, torque ${\bf T}^{\prime}$, force ${\bf F}^{\prime}$, fluid velocity field ${\bf v}^{\prime}$, and isotropic fluid stress ${\bm \sigma}^{\prime}$), defined on the same exterior domain of the swimmer $D$, as follows
\begin{equation}
   \Omega'_i T_i + U'_i F_i - \Omega_i T'_i - U_i F'_i = \int_{\partial D} {u_S}_i \sigma'_{ij} \nu_j \dd{S} -  \int_D \Sigma_{ij} \partial_i v'_j \dd{V}.
   \label{eq:Lorentz_main}
\end{equation}
The general boundary condition on the particle in the complex fluid is $\mathbf{v}\big|_{\partial D}
  = \mathbf{U} + \boldsymbol{\Omega}\times\mathbf{x} +
  \mathbf{u}_S,$
where $\mathbf{u}_S$ is a velocity on the swimmer's surface $\partial D$ with normal $\bm{\nu}$. The right-hand side of Eq. \ref{eq:Lorentz_main} includes a surface integral coming from the boundary condition in the isotropic fluid~\cite{stone1996propulsion} and a volume integral coming from the complex fluid stress. We use the standard Ericksen-Leslie formulation $\Sigma_{ij}=\Sigma^v_{ij}+\Sigma^\mathrm{el}_{ij}$ for the viscous stress (Eq. \ref{s_viscous}) and elastic stress (Eq. \ref{s_Er}). To obtain an analytic solution and resolve the anisotropic term, we take the lowest order of anisotropic viscosities compared to the isotropic viscosity, and expand the out of equilibrium driven director solution perturbatively up to second order in terms of equilibrium multipolar moments (dipolar moment $\alpha$, quadrupolar moment $q$) and $\mathrm{Er}$ as shown in Eq.~\ref{dir_expansion}.

For the propulsion velocity of a rotating sphere, Eq.~\ref{eq:Lorentz_main} simplifies significantly as the surface integral term, which is the usual driving force for propulsion in Newtonian fluids, is identically zero.  Eq.~\ref{eq:Lorentz_main} simplifies to
$  U_i F'_i =    \int_D \Sigma_{ij} \partial_i v'_j \dd{V}$. This form reveals that the anisotropic stress $\vb{\Sigma}$, which contains information about the director field and the flow field, plays a central role in determining the sphere's motion. Since $\vb{\Sigma}$  is linear in driving frequency $\mathit\Omega$ and the equilibrium director field already has broken symmetry, the propulsion velocity is also linear in $\mathit\Omega$. Evaluation of the term $\int_D \Sigma_{ij} \partial_i v'_j \dd{V}$ within our perturbation analysis gives the expression for the propulsion velocity
\begin{align}
    \vb{U} = \begin{bmatrix}
        U_x \\ U_y 
    \end{bmatrix}=\text{Er} \begin{bmatrix}
         \alpha\, f_1(\lambda)+ \alpha q \,f_2(\lambda) \\  \text{Er}  \alpha \,f_3(\lambda) 
    \end{bmatrix}\frac{K}{\mu R},
    \label{eq:propulsion}
\end{align}
where $f_1(\lambda)$, $f_2(\lambda)$ and $f_3(\lambda)$ are functions specific for the swimmer's shape and boundary conditions (see Methods). These functions depend on nematic viscosity parameters, which we can compactly reduce into one alignment parameter $\lambda$ as explained in the Methods section.
For spheres rotating in a nematic fluid, we predict that the propulsion velocity $U_x$ is linear in  $\mathit\Omega$, included in the Ericksen number $\text{Er}=\mathit\Omega R^2 \gamma/K$.
Furthermore, flow-induced perturbations in the director field lead to motion along $y$-direction. We perform the calculation for the spherical homeotropic swimmer, but a similar approach can in principle be carried out for arbitrary shaped objects.

Theory predicts that $U_x \propto \mathrm{Er}$ for the rotating sphere with the swimming velocity induced by the dipolar and quadrupolar contributions to the equilibrium structure. Using nematic material parameters and viscosity dependent functions $f_1$, $f_2$ and $f_3$, we estimate the swimming direction and amplitude for different nematic environments (see Methods). For 5CB, theory agrees well with the experiment (Fig.~\ref{figure 2}E) where fitting the data indicates $U_x \propto \mathrm{Er}^{1.02\pm 0.05}$. The theory also indicates that swimming is predominantly in the direction $\boldsymbol{\Omega}\times\vb{p}$ (Fig.~\ref{figure 2}C,F), where $\vb{p}$ is the dipolar vector from colloid to the hedgehog and $\boldsymbol{\Omega}$ is the colloidal rotation vector. Along the director field, the swimming velocity $U_y$ depends on flow induced director distortions and is quadratically dependent on $\text{Er}$.

\begin{figure} [b!] 
	\centering
\includegraphics[width=0.66\textwidth]{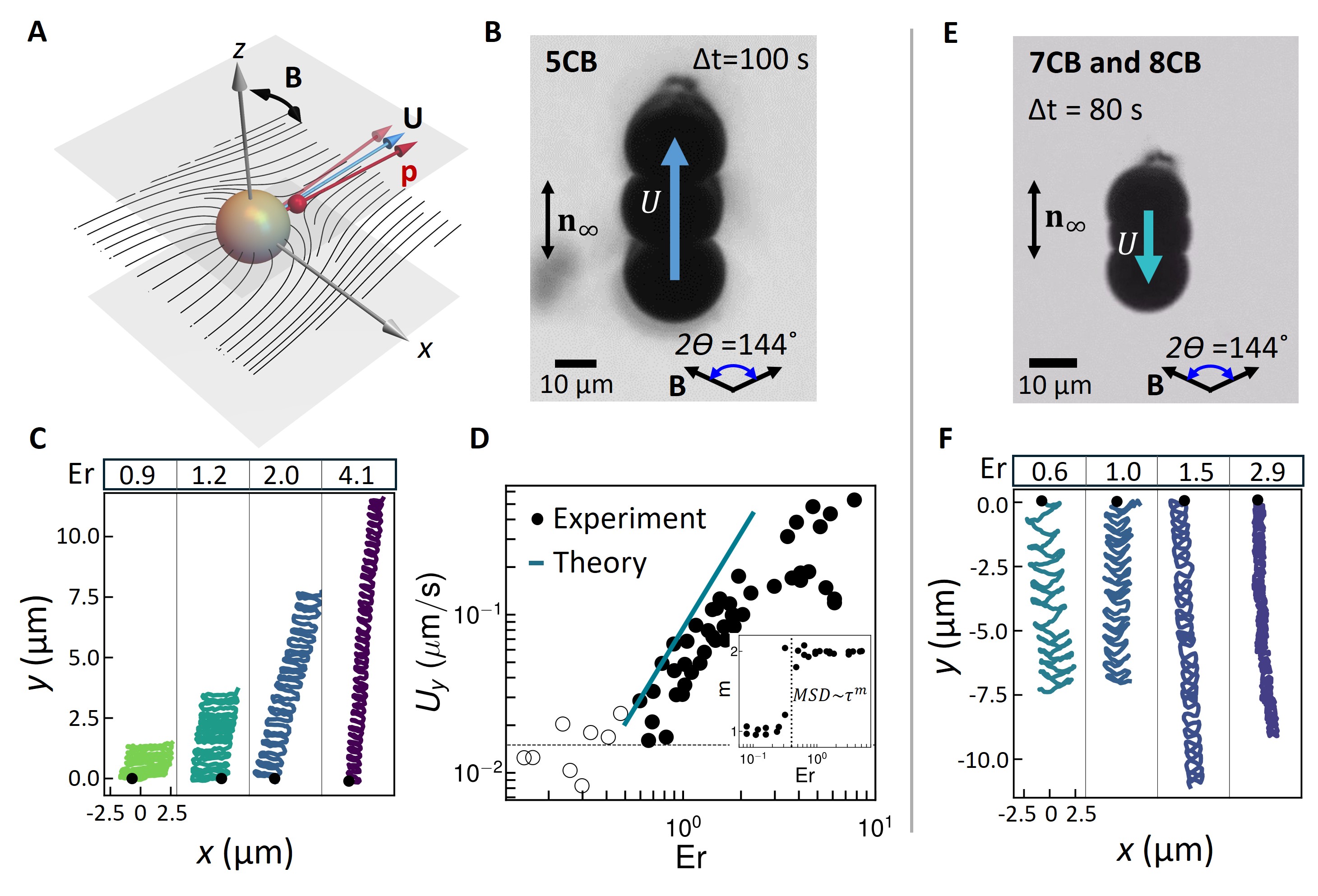} 
	\caption{\textbf{~Swimming of spherical colloids with a reciprocal driving force in nematic liquid crystals.}
		 (\textbf{A}) Schematic of a reciprocal forcing of a sphere placed in a nematic liquid crystal. (\textbf{B}) Overlay of sphere textures during swimming in 5CB along the far field director driven by a reciprocal forcing for $\mathrm{Er}=3.8$ ($\mathit\Omega=3.14~s^{-1}$). The arrow represents the swimming trajectory. $\Delta t$ is the time interval between two consecutive overlaid images. (\textbf{C}) Sphere trajectories in 5CB at different $\mathrm{Er}$ over elapsed time $t=40~s$. (\textbf{D}) Swimming speed along the far field director as a function of $\mathrm{Er}$. The open circles represent the limit of $\mathrm{Er}$ in which the random Brownian motion dominates the directional swimming. The solid blue line represents the theoretical prediction, which has a slope of 2. The inset is the exponent of $\mathrm{MSD}$ vs $\mathrm{Er}$, which sets the limit on $\mathrm{Er}$ below which the motion is Brownian. All material parameters used to plot the theoretical dependence are available in the Methods section. (\textbf{E}) Overlay of sphere images during swimming in the mixture of 7CB and 8CB along the director driven by a reciprocal forcing for $\mathrm{Er}=2.0$ ($\mathit\Omega=3.14~s^{-1}$). The arrow represents the swimming trajectory. $\Delta t$ is the time interval between two consecutive overlaid images. (\textbf{F}) Sphere trajectories in the mixture of 7CB and 8CB at different $\mathrm{Er}$ over duration  $t=40~s$.
	}
	\label{figure 3} 
\end{figure}

\subsection*{Reciprocal swimming strokes}
Microswimmers in Newtonian fluids generally move at very low Reynolds numbers, since the Reynolds number scales with the characteristic size of the swimmer. Consequently, in isotropic fluids, reciprocal swimming strokes are unable to produce net locomotion.
However, our observation of translation of the rotating spheres demonstrates that stresses in structured fluids differ significantly from those in isotropic fluids. Analysis shows that the swimming speed relies on the degree to which viscous stresses exerted by the swimmer's motion disrupt the underlying director field, as characterized by the Ericksen number. This motivates us to explore whether reciprocal forcing can drive flow, i.e. to probe whether a sphere which performs a reciprocal swimming stroke, by rotating first in one direction over some angle, and then in the opposite direction, can swim.

Analysis shows that for such a reciprocal swimming stroke, the propulsion terms linear in $\mathrm{Er}$ average out to zero as they are proportional to $\sin(\mathit\Omega t)$ and the integrand structure is odd under time reversal. Only the swimming velocity component parallel to the far field director $\mathbf {n_\infty}$ remains due to nonlinearity arising through a combination of the equilibrium structure in the surrounding fluid and rate dependent correction proportional to $\mathrm{Er}$. The velocity is proportional to $\sin^2(\mathit\Omega t)$, which results in survival of the component proportional to the Ericksen number squared $U \propto \mathrm{Er}^2$ (SI) and is enabled through nonlinear breaking of the time-reversal symmetry in the viscous stress, while the spatially broken symmetries are provided by the surrounding complex fluid along the $x$- and $y-$directions.

In experiment, to generate a reciprocal swimming stroke, we impose a magnetic field $B$ that rotates our sphere between $-\theta$ and $\theta$ at angular velocity $\mathit\Omega$ around the $z$-axis, alternating between CW and CCW rotation (Fig.~\ref{figure 3}A). The disturbance in the nematic director field is apparent in the periodic displacement of the hedgehog defect. Under this reciprocal forcing, the colloids translate roughly parallel to $\mathbf {n_\infty}$ while the defect leads the way~(Fig.~\ref{figure 3}B, C and Movie S3) as predicted by the theory. The swimming speed is more than an order of magnitude slower than the continuously rotated sphere. To probe the swimming speed's dependence, the Brownian motion of the microscale colloids must again be addressed  at low $\mathrm{Er}$. We use the observed particle trajectories to construct the mean squared displacement ($\text{MSD}$) versus lag times $\tau$ (Fig. \ref{Sup 5CB Oscillation}). For particles undergoing Brownian displacement, $\text{MSD}\sim \tau$ whereas for particles moving ballistically with a uniform speed $\text{MSD}\sim \tau^2$. The values of the exponent $\text{m}$ of the $\text{MSD}$ versus $\mathrm{Er}$ are summarized in Fig.~\ref{figure 3}D Inset. Under very slow rotation ($\mathrm{Er}<0.6$) the diffusive displacement is larger than the displacement by swimming (Fig.~\ref{figure 3}D). At $\mathrm{Er}=0.6$, the swimming and diffusive displacements are similar, and for $0.6<\mathrm{Er}<1.6$, swimming dominates the particle's motion (Fig.~\ref{figure 3}D). In this regime, fitting the data indicates $U_y \propto\mathrm{Er}^{1.85\pm 0.30}$, which is in rough agreement with the quadratic  dependence predicted by theory (Fig.~\ref{figure 3}D). Theory also predicts that the velocity of spherical microswimmers depends on the LC's viscoelastic properties, characterized by the alignment parameter $\lambda$. Thus, altering $\lambda$ is expected to modify both the speed and trajectory of the microswimmers (Fig.~\ref{fig:swimming-direction}). All of the experiments presented thus far were performed for colloidal particles immersed in the nematic liquid crystal 5CB, for which $\lambda>1$.
We confirmed this prediction by replacing 5CB with a 1:1 mixture of 7CB and 8CB, which has $\lambda<1$~\cite{Ternet1999}. In this environment, the continuously rotated sphere moves strongly in the $x$-direction and weakly in the $-y$-direction (Fig. \ref{Sup 7CB Rotation} and Movie S4). Reciprocal rotation of spheres results in a translation in the opposite direction of the reciprocally forced spheres in 5CB; here the sphere leads the way and the defect follows (Fig.~\ref{figure 3}E,F and Movie S5).

It is natural to ask whether the bounding walls play an important role in the observed motion of the rotated sphere. Anchoring
at the walls defines the far field nematic director and allows the companion defect to
form. Furthermore, the walls provide effective elastic repulsion, giving a surface against which gravity
and elasticity set an equilibrium height for the colloid below the midplane of the
cell. The walls also screen the range of the nematic distortion and contribute to the
decay of the flow field generated by sphere rotation. Thus, the flow
and director fields are three dimensional with weak top-bottom asymmetry, and this
asymmetry could generate additional stress tensor components that alter both the
speed and direction of motion. Overall, the agreement between theory based on an unbounded domain and experiment suggests that these effects are not dominant in our current experiments.

\begin{figure}[t!]
    \centering
    \includegraphics[width=0.66\linewidth]{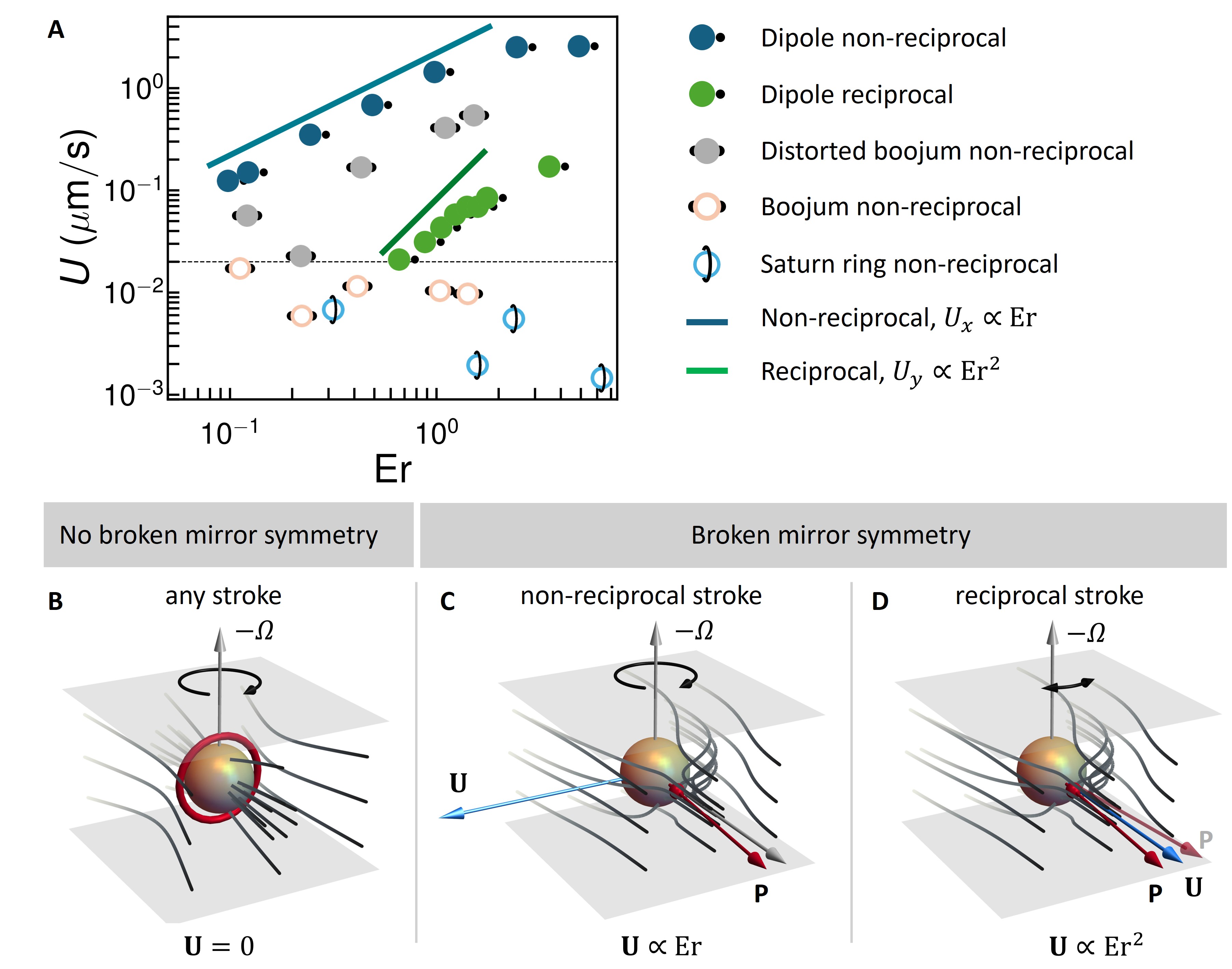}
    \caption{ 
    \textbf{~Demonstration of Scallop theorem for anisotropic fluids.}
    {\bf (A)} Swimming speed as a function of $\mathrm{Er}$ for spheres accompanied with a point defect, Saturn ring, or boojums under reciprocal or non-reciprocal strokes. The solid lines indicate the linear dependence for non-reciprocal (blue) and quadratic dependence for reciprocal (green) forcing. {\bf (B)}-{\bf (D)} Delineation of swimming strokes in structured fluids: {\bf (B)} Configurations that do not break the reflection symmetry, for example a sphere creating quadrupolar distortion like a Saturn defect ring. Such an object cannot swim upon rotation or oscillation in the magnetic field. Without broken symmetries, translation cannot be generated. {\bf (C)} Configurations that break reflection symmetry, for example a sphere creating a dipolar distortion with a point defect. Upon performing a non-reciprocal swimming stroke, in this case rotation, the sphere moves in the direction of $\vb{U} \propto {\boldsymbol\Omega} \times \vb{p}$. Swimming speed scales linearly with the Ericksen number. {\bf (D)} Configurations with broken reflection symmetry, e.g. the sphere with dipolar director configuration. Reciprocal swimming strokes propel the swimmer. The swimming speed scales quadratically with the Ericksen number. In our example, such a stroke can be achieved via oscillating magnetic field. The sphere moves parallel to the director far field $\vb{U} \parallel \vb{n_\infty}$.}
    \label{figure 4}
\end{figure}

\subsection*{Scallop Theorem in anisotropic fluids}
For microscopic swimmers, moving in structured fluids differs fundamentally from swimming in an isotropic fluid, which requires highly specialized structures like flagella to break time-reciprocity and spatial symmetries. In nematic environments, the fluid’s molecular ordering provides a distinct mechanism to break these constraints, obviating the need for complex apparatus and swimming strokes. To demonstrate this idea, we presented the example of continuous axisymmetric rotation of a sphere with a dipolar nematic field.  The sphere’s rotation distorts the nematic field, causing it to move nearly perpendicular to the far field nematic director at speeds proportional to the rotation rate $\mathit\Omega $ (Fig. \ref{figure 4}A,C).  We further show that reciprocal forcing of the sphere also generates translation with swim speeds that are quadratic in  $\mathit\Omega $ (Fig. \ref{figure 4}A,D).  
In these examples, the spheres' symmetric motions generate broken symmetries in the nematic liquid crystal. Theory also predicts that highly symmetric configurations will not generate the required broken symmetries to enable propulsion. For example, a rotated sphere with a Saturn ring in a planar cell moves in a manner indistinguishable from weak drift (Figs. \ref{figure 4}(A,B),\ref{sup saturn ring} and Movie S6). Spheres with tangential anchoring form surface defects known as boojums and exhibit a symmetric equilibrium configuration. Under rotation, these spheres show two distinct behaviors. Spheres that weakly distort the surrounding nematic remain effectively stationary, with speeds comparable to the background drift (Fig. \ref{figure 4}A and Fig. \ref{sup boojum}). In contrast, spheres that induce strong director distortions exhibit complex defect dynamics, including deformation of the boojums that can break symmetry and lead to translational motion (Fig. \ref{figure 4}A and Fig. \ref{sup boojum}).

We extend our theoretical analysis to show that the same scaling of propulsion velocity with the stroke rate
is expected also for swimmers of arbitrary shapes beyond the spherical colloids. The net displacement of the swimmer into the $k$-th direction $d_k$ may be decomposed as displacement due to standard isotropic and novel anisotropic contributions $d_k 
    =
    d_k^{\mathrm{iso}}
    +
    d_k^{\mathrm{ani}},$
with
$
d_k^{\mathrm{iso}}
=
-
\int_0^T M_k
\int_{\partial D}
{u_S}_i
{\sigma}'^{(k)}_{ij}
\nu_j
\dd{S} \dd{t}
$
and $d_k^{\mathrm{ani}}
=
\int_0^T
M_k
\int_{D}
\Sigma_{ij}
\partial_i v_j'^{(k)}
\dd{V}
\dd{t},$ where the definitions include solutions for the isotropic stress tensor $\bm{\sigma}'$ and flow $\vb{v}'$ of the auxiliary problem combined with  mobility elements for the $k$-th direction $M_k$. The condition for the net propulsion ($d_k\ne0$), is determined by whether the integrands (which depend on the swimmer's shape, the fluid flow and the fluid's structure) are odd under the time reversal operator and if they break any spatial mirror symmetries. For the isotropic part, as by Purcell's argument, the swimmer has to perform nonreciprocal strokes and additionally break spatial reflection symmetry in order to propel \cite{purcell2014life,Lauga2009}. However, in anisotropic fluids, time-reversal symmetry can also be broken by nonlinear couplings which enter via the additional stress $\Sigma_{ij}$. To generalise, therefore, symmetry breaking is not  determined solely by  the swimmer's shape, but must be considered as a combination of both the swimmer and the structure of complex fluid.

In Newtonian fluids, swimming strokes that generate reciprocal distortions do not generate propulsion, even if parts of the stroke occur at different rates. In complex fluids, nonlinear couplings, for example of director structure and flow, lead to a propulsion velocity for reciprocal swimmers that is at least quadratic with the Ericksen number. In nematic fluids, the Ericksen number can exceed 1 even for microscale objects; therefore, despite the quadratic scaling with Er, reciprocal swimmers can still effectively propel and escape the standard notion of Purcell’s theorem for isotropic fluids. For example, for 5CB, $\mathrm{Er}=3\cdot10^5\,\mathrm{Re}$, indicating that flow induced changes to the director field can be significant deep in the creeping flow regime.

Strategies to develop functional materials and responsive active materials are of driving interest to materials scientists and engineers.  To provide general guidelines on propulsion in anisotropic environments, we reformulate the Scallop theorem for nematic fluids.
We show that any swimming stroke that generates non-symmetric distortions in the nematic field will be effective for microscale swimmers. 
The Scallop Theorem in nematic fluids categorizes swimmers in three ways.
First are swimmers that break time-reciprocity with their stroke and would also swim in isotropic fluids; bacteria~\cite{shiyanovskii2005lyotropic,mushenheim2014dynamic,peng2016command}
are a prominent example. Bacteria that swim effectively in isotropic fluids also move effectively in nematic fluids, where their trajectories and interactions are influenced by the nematic field.  Second are swimmers that generate broken symmetries in the nematic liquid crystal and move under continuous non-reciprocal forcing.  These include our example of the continuously rotated sphere and the examples of swimming rotated disks~\cite{yao2022topological} and microrobots~\cite{yao2022nematic} in the literature. Third are swimmers that move under reciprocal forcing, which also exploit the nematic’s broken symmetries.  One example is our reciprocally forced sphere. Another is a pulsating bubble in a dipolar configuration, which moves under the action of different stresses fore and aft the bubble as its companion hedgehog is displaced periodically by bubble pulsation  \cite{kim2024symmetrically}. 
The swimming velocity for non-reciprocal swimmers of type one and two scales linearly with the stroke rate. The swimming velocity of reciprocal type three swimmers scales quadratically with the stroke rate. 

\vspace{0.4 cm}
\noindent

\section*{Discussion}
In the paper, we show how fluid anisotropy and shear-dependent structural response result in novel propulsion modes in nematic fluids and provide rules on scaling of the propulsion speed for reciprocal and non-reciprocal swimming strokes with the dimensionless Ericksen number. We present a minimal realization of a nematic microswimmer, a rotating sphere, for which both experimental observations and theoretical predictions are in quantitative agreement.
We show that due to inherent broken symmetries around colloidal inclusions in nematic fluids, even a continuously rotating sphere propels with propulsion velocity linearly dependent on the rotation rate. Such propulsion occurs even for negligible flow-induced director distortions. Director-flow coupling becomes important for reciprocal sphere rotation, where propulsion velocity is determined to be quadratically dependent on the rotation rate. 

The breaking of the Scallop Theorem through nonlinear (quadratic) effects is universal for complex fluids and has been reported, for example, in viscoelastic media, where the nonlinearity is characterized with Deborah instead of Ericksen number. However, there the main mechanism for temporal symmetry breaking arises through rate dependent memory signature in the stress ~\cite{yasuda2020reciprocal,datt2018two,kroo2022freely,puente2019viscoelastic,eberhard2023reciprocal} rather than through structure. For this reason, nematic and polar fluids exhibit a far more diverse set of effective swimming strokes. 

The director--flow coupling in nematics enables propulsion mechanisms that are fundamentally distinct from those identified in conventional viscoelastic fluids. In particular, the spontaneous emergence of complex director configurations can break reflection symmetry even in equilibrium, eliminating the need for geometrically complex swimmers such as scallops or multi-body structures. At low Reynolds numbers, traditional swimmers require multiple degrees of freedom, such as the hinges used in the original paper by Purcell~\cite{purcell2014life}. Likewise, for pulsating bubbles in nematic fluids~\cite{kim2024symmetrically}, reciprocal forcing alone is insufficient for propulsion: the swimming stroke requires an additional dynamical degree of freedom supplied by the fluid, namely a phase lag between the bubble pulsation and the director response. Consequently, no propulsion occurs in the quasi-static limit, where the phase lag vanishes and time-reversal symmetry is restored. In contrast, we demonstrate that rotating spheres do not require any additional degrees of freedom. The equilibrium director configuration prescribed by the swimmer inclusion, combined with the flow-induced perturbation generated by rotation (i.e. the rate of strain or co-rotational derivative of the director), is sufficient to produce net propulsion. 

Moreover, propulsion needs not rely solely on mechanical actuation: it can be driven or modulated via external fields that couple to the director, including electric fields, time-dependent surface anchoring, or ionic currents.

Within the context of flow-director coupling, several liquid-crystal-enabled microswimmers have been constructed in the past, which can be sorted according to the general rules developed in the article classified as reciprocal~\cite{senyuk2024out,eremin2015optically, kim2024symmetrically} or nonreciprocal~\cite{yao2022topological,yao2022nematic,zhang_sweeping_2026} swimmers. The analytical framework also provides guidelines for further development of new microswimmers and for the theoretical understanding of their motion. 

Our analysis can be extended to a broad range of swimming modes. In this work, we focus on spherical swimmers that are force-free, but not torque-free. The same framework is readily applicable to torque-free swimmers, including minimal models such as counter-rotating dumbbells and squirmers, and more generally to other classes of microswimmers in liquid crystals~\cite{yao2022topological,yao2022nematic,rajabi2021directional,kim2024symmetrically,eremin2015optically,tavera2025quorum,lintuvuori2017hydrodynamics,mandal2021multiparticle,senyuk2024out,Lavrentovich2010,hernandez2014reconfigurable}. The only essential requirement is the availability of sufficiently accurate analytical or approximate solutions for the flow and director fields generated by the swimmer, including time-dependent solutions where necessary. In the future, it would be of interest to apply similar methodology to describe liquid crystal microswimmers driven by random flows~\cite{rajabi2021directional} or electrophoresis~\cite{Lavrentovich2010,hernandez2014reconfigurable}.

In conclusion, in structured fluids even the simplest swimmer geometries like isotropic spheres effectively propel. Therefore, our experimental and theoretical framework provides a foundation for investigating collective effects and dense suspensions
without requiring intricate fabrication or control strategies. This opens a pathway toward scalable implementations and, ultimately, the design of novel functional and responsive active materials. We demonstrate the effectiveness of such swimming strategies in liquid crystals, but similar mechanisms should also be relevant for non-equilibrium biological colloidal dynamics in anisotropic environments in nature. Essentially, rather than imposing additional constraints, the structure of the fluid relieves the requirement for complex swimming strokes, and almost anything goes.

\newpage
\clearpage

\bibliographystyle{unsrtnat}  
\bibliography{references}  

\newpage
\clearpage

\section*{Methods}
\section*{Experimental setup}
NLC-filled cells are prepared from two microscope slides separated by 50 µm gap defined by particle spacers that are fixed by a UV polymerizable glue (NOA 68). The inner surfaces of the glass substrates are coated with polyimide PI2555 (HD MicroSystems) and rubbed unidirectionally to impose planar alignment on the nematic director. The cells are filled with either 4-cyano-4'-pentylbiphenyl (5CB) or a 1:1 mixture of 4-Cyano-4'-heptylbiphenyl (7CB) and 4-Cyano-4'-octylbiphenyl (8CB). Ferromagnetic nickel coated glass spheres (Cospheric) are coated with N-dimethyl-N-octadecyl-3aminopropyl-trimethylsilyl chloride (DMOAP) to impose perpendicular (homeotropic) anchoring of the liquid crystal molecules.  
The spheres are dispersed in NLC, injected into the cell and allowed to equilibrate in the nematic phase. Two pairs of Helmholtz coils are arranged to generate a uniform rotating magnetic field B $\sim$ 10 mT at the center of the domain, well below field strengths that alter NLC orientation. The spheres’ motion is observed using a Zeiss Axio Vert A1 inverted microscope and recorded by a FLIR camera. Trajectories are tracked via a Python code.

\section*{Parameters}
The following parameter values were used to plot the experimental and theoretical results:
sphere radius $R=11 $~$\upmu$m, nematic average elastic constant $K=4 $~pN, rotational viscosity $\gamma_1$ = 30 ~mPa s, isotropic viscosity $\mu$=37 ~mPa\,s, and alignment parameter $\lambda$=1.05. For the  dipolar and quadrupolar moments of the equilibrium director field around the sphere, we use $\alpha$=3.08 and $q$ = 1.5, following~\cite{stark_physics_2001}. The ratio between Reynolds number $\mathrm{Re}=\frac{\rho\Omega R^2}{\mu}$  and Ericksen number $\mathrm{Er}=\frac{\gamma_1\Omega R^2}{K}$ equals $\mathrm{Er}=\frac{\gamma_1\mu}{\rho K}\mathrm{Re}$, which for the above parameters and nematic density $\rho=10^3\,\text{kg/m}^3$ gives $\mathrm{Er}=3\cdot10^5\,\mathrm{Re}$.

\section*{Propulsion velocity of spherical swimmer}
To determine the propulsion velocity, we employ a method that relates two Stokes flow problems defined on the same domain: the physical problem of interest and an auxiliary problem with known solutions. This approach enables the use of classical results from Newtonian hydrodynamics, such as analytical flow fields and drag coefficients, to address the more complex problem of a self-propelled swimmer. The connection between the two systems is established through an integral identity known as the Lorentz reciprocal theorem \cite{Lauga2009}. Importantly, this theorem can be generalized to arbitrary complex fluids \cite{neville2024controlling}, providing a versatile framework for analyzing microswimming in nematic environments. By considering a rotating spherical swimmer, isotropic propulsion mechanisms are eliminated, allowing us to isolate the contributions arising solely from the anisotropic nematic medium. Consequently, this system represents a minimal swimmer whose locomotion is entirely mediated by nematic anisotropy.

\subsection*{Lorentz theorem for complex fluids}
We define our calculation domain as $D=\{(x,y,z)\in  \mathbb{R}^3 \, | \, x^2+y^2+z^2 \ge R^2 \}$, which represents the bulk around the spherical colloid. In our case, we choose the auxiliary problem of a sphere dragged through an isotropic fluid by an external force $\vb{F'}$ with no applied torque and no angular velocity $\vb{\Omega'}$. On the other hand, the real sphere is experiencing an applied torque $\vb{T}$ and has an angular velocity $\vb{\Omega}$, but it is force free. Using these assumptions we end up with the following relation
\begin{equation}
   \vb{U} \cdot \vb{F'} =  \int_D \Sigma_{ij} \partial_i v'_j \dd{V}.
   \label{eq:Lorentz-1}
\end{equation}
We use well-known solutions for the auxiliary (isotropic) problem: The auxiliary force is given by the Stokes' law $\vb{F'}= -6\pi \mu R \vb{U'} $ and the corresponding bulk auxiliary velocity field by $v'_i (\vb{r})= \frac{1}{4} \frac{R}{r} (3+\frac{R^2}{r^2}) V'_i + \frac{3}{4} \frac{R}{r} (1-\frac{R^2}{r^2})\frac{x_i x_j}{r^2} V'_j$. Applying Stokes's law to Eq.~\ref{eq:Lorentz-1}, we obtain
\begin{equation}
    \vb{U} \cdot \vb{U'} = -\frac{1}{6\pi \mu R} \int_D \Sigma_{ij} \partial_i v'_j \dd{V}.
    \label{eq:Lorentz}
\end{equation}
Eq.~\ref{eq:Lorentz} gives a nontrivial relation between the propulsion velocity of the sphere in a complex fluid and a stress tensor that captures a response of the fluid. The difficulty arises in finding a solution for the bulk stress tensor $\Sigma_{ij}$, which depends on the coupled solutions of the Stokes' equation for the flow field and the evolution of the order parameter.
\\ \\
\subsection*{Director field perturbation}
We start by writing a multiple-scale expansion of the director with the smallness parameter $\delta$ characterized by the main dimensionless parameters of the system
\begin{equation}
    \delta \sim \alpha \sim q \sim \mathrm{Er} \ll 1,
\end{equation}
where $\alpha$ is the equilibrium dipole magnitude, $q$ is the equilibrium quadrupole magnitude and $\mathrm{Er}=\frac{\gamma_1 \Omega R^2}{K}$ the Ericksen number of the rotating sphere.
We expand the director field around a homogeneous director $\vb{n_0}$ with corrections $\delta^n \vb{n_n} \perp \vb{n_0}$
\begin{equation}
    \vb{n} = \vb{n_0} + \delta \vb{n_1} + \delta^2 \vb{n_2} + ... +\delta^n \vb{n_n} +...,
\end{equation}
where we define $\delta^n$ as a multilinear form operator of $\alpha$ and $\mathrm{Er}$ eg. $\delta^2 \vb{n_2}= \alpha^2 \vb{n^{(1)}_2}+ \alpha \mathrm{Er} \vb{n^{(2)}_2} + \mathrm{Er}^2 \vb{n^{(3)}_2}$.
\\ \\ 
However, by definition, the director has to be normalized such that $|\vb{n}|=1$. Taking into account the normalization, we find
\begin{equation}
    \vb{n} =\frac{\vb{n_0} + \delta \vb{n_1} + \delta^2 \vb{n_2} + ... \delta^n \vb{n_n}+...}{|\vb{n_0} + \delta \vb{n_1} + \delta^2 \vb{n_2} + ... \delta^n \vb{n_n}+...|} \approx \vb{n_0} + \delta \vb{n_1} + \delta^2 \vb{n_2} - \vb{n_0} \frac{1}{2} |\delta \vb{n_1}|^2  + \mathcal{O}(\delta^3).
    \label{dir_expansion}
\end{equation}
We use a flow field generated by a rotating sphere expanded in the zeroth order of the viscous anisotropy (six nematic viscosities are smaller than the isotropic viscosity $\alpha_i\ll\mu \, \,\,  \forall i \in \{1,... ,6\}$), which can be written for rotation around $z$-axis as $\vb{\delta v}(\vb{r})=\Omega R^3 \frac{(y,-x,0)}{r^3}=\frac{\text{Er} K R}{\gamma_1} \frac{(y,-x,0)}{r^3}$. 

\subsection*{Director field around rotating sphere}

The dynamics of the director field $\vb{n}(\vb{r},t)$ is captured by the Ericksen-Leslie formalism describing advection, flow-induced reorientation, and elastic relaxation towards equilibrium
\begin{equation}
\frac{K}{\gamma_1} \left(\nabla^2 \vb{n} -\vb{n} (\vb{n}\cdot\nabla^2 \vb{n})\right)=
    \partial_t \vb{n} + \vb{v} \cdot \nabla \vb{n}  +\vb{\Omega} \vb{n} 
    +\frac{\gamma_2}{\gamma_1} \left( \vb{A} \vb{n} - \vb{n} (\vb{n} \cdot \vb{A} \vb{n})\right), \label{director-Eq}
\end{equation}
where $A_{ij}=\frac{1}{2}(\partial_i v_j + \partial_j v_i)$, $\Omega_{ij}=\frac{1}{2}(\partial_i v_j - \partial_j v_i)$, and for stationary states that we consider in the paper to compute the propulsion of rotating spheres $\partial_t\vb{n}=0$.
\\ \\
\noindent We first calculate the director equation up to the first order, where we collect only terms of order $\mathcal{O}(\delta)$
\begin{equation}
    \frac{K}{\gamma_1} \nabla^2 \delta \vb{ n_1} = \delta\vb{ \Omega} \vb{n_0} + \frac{\gamma_2}{\gamma_1} ( \delta \vb{ A} \vb{n_0} - \vb{n_0} ( \vb{n_0}\cdot \delta \vb{A}\vb{n_0})) + \mathcal{O}(\delta^2). \label{first-order-equation}
\end{equation}
By collecting second order terms $\mathcal{O}(\delta^2)$, and using the lower order equation we arrive at
\begin{align}
    \frac{K}{\gamma_1} \nabla^2 \delta^2 \vb{ n_2} =&  \delta \vb{v} \cdot \nabla \delta\vb{ n_1} + \vb{\delta \Omega}\delta \vb{ n_1} + \vb{n_0} (\delta \vb{n_1} \cdot \delta \vb{\Omega} \vb{n_0}) \nonumber
    \\
    &+ \frac{\gamma_2}{\gamma_1} (\delta \vb{ A} \delta \vb{ n_1}  - \vb{n_0} (\vb{n_0} \cdot \vb{\delta A} \delta \vb{n_1}) - \delta \vb{n_1} (\vb{n_0}\cdot \delta A \vb{n_0}) ) + \mathcal{O}(\delta^3). \label{second-order-equation}
\end{align}

Let us first define the far field as $\vb{n_0}=\vb{e_y}$. 
The first order equation for the director deviations $\delta \vb{ n_1}$ (Eq.~\ref{first-order-equation}) has a homogeneous solution---the multipole expansion of the director field in equilibrium---plus a particular solution
\begin{equation}
    \delta \vb{n_1} =\vb{n_1}^{\mathrm{dip.}} + \vb{n_1}^{\mathrm{quad.}} + \vb{n_1}^{\mathrm{part.}} =\alpha \vb{\tilde{n}_1}^{\mathrm{dip.}} + q \vb{\tilde{n}_1}^{\mathrm{quad.}} + \mathrm{Er}\vb{\tilde{n}_1}^{\mathrm{part.}},
\end{equation}
where we expressed the structure by explicitly taking the main dimensionless parameters out from different contributions. We now take a closer look at an actual structure of the contributions.
\\ \\
The dipoles, harmonic solutions of the Laplace equation with $l=1$, are spanned in a vector space which consists of two real-dimensional vector spaces for each spin $s=\{0,1,2\}$ \cite{houston_active_2023}. In our case we will focus on an equilibrium UPenn dipole of the type
\begin{equation}
    \vb{n_1}^{\mathrm{dip.}} = \frac{\alpha R^2}{r^3} (x,0,z).
    \label{eq:equilibrium_dipole}
\end{equation}
In addition we consider also a quadrupolar contribution
\begin{equation}
    \vb{n_1}^{\mathrm{quad.}} = \frac{q R^3}{r^5} \left( y x, 0, y z  \right).
    \label{eq:equilibrium_quadrupole}
\end{equation}
The particular solution is obtained from the Poisson equation of the form
\begin{align}
    \nabla^2 \bm{n_1}^{\mathrm{part.}} =  \frac{ \gamma_1 R^3 \Omega}{4 K r^3}\begin{bmatrix}
      -\left(-6 \lambda  \sin ^2(\theta ) \cos (2 \phi )+3 \cos (2 \theta )+1\right)
    \\
    0
    \\
    3 (\lambda +1) \sin (2 \theta ) \cos (\phi )
    \end{bmatrix}.
\end{align}
The equation above in the generic form $ \nabla^2 \vb{n}^{\mathrm{part.}}(r,\theta,\phi)=\frac{\vb{\rho} (\theta,\phi)}{r^3}$ can be solved by using the Green's function for Poisson equation in spherical coordinates as $$\vb{n}^{\mathrm{part.}}(r,\theta,\phi)=-\frac{1}{r}\sum_{l=1}^{\infty}\sum_{m=-l}^{l}\frac{Y_{l,m}}{l(l+1)}\bra{\vb{\rho}(\theta, \phi)}\ket{\mathrm{Y}_{l,m}(\theta, \phi)}^*,$$ where $\braket{f(\theta, \phi)}{g(\theta, \phi)}=\oint f(\theta, \phi) g(\theta, \phi) \sin\theta \dd{\theta} \dd{\phi}$.
\\ \\
For the components of $\vb{n}^{\mathrm{part.}}$, we get contributions from even $l$ with decaying magnitude as $l$ increases. We note that only $l=2$ term contributes in the zeroth anisotropy regime. We obtain
\begin{align}
    \vb{n_1}^{\mathrm{part.}} = &\frac{\gamma_1 \pi R^3 \Omega }{4 K r}
    \begin{bmatrix}
        - \lambda \cos (2 \phi) \sin ^2(\theta)+ \cos ^2(\theta)-\frac{1}{3}
        \\
        0
        \\
        -\frac{(\lambda+1)}{2}  \cos (\phi) \sin (2 \theta) 
    \end{bmatrix}.
\end{align}
We introduced Ericksen number as $\mathrm{Er}= \gamma_1 \Omega R^2/K$. Combining the particular solution and the equilibrium  dipolar plus quadrupolar solutions $\delta \vb{n^{\text{eq.}}_1}
    = \frac{\alpha R^2}{r^2} (\cos \phi \sin\theta\vb{e_x}+ \cos\theta \vb{e_z}) + \frac{q R^3}{r^3} (\sin^2 \theta \sin 2\phi \vb{e_x} + \sin 2\theta \sin \phi \vb{e_z})$, we obtain 
\begin{align}
    \delta \bm{n_1} = & \frac{\alpha R^2}{r^2} (\Tilde{Y}_{1,1}\vb{e_x}+ \Tilde{Y}_{1,0} \vb{e_z}) + \frac{q R^3}{r^3} ( \Tilde{Y}_{2,-2} \vb{e_x} +  \Tilde{Y}_{2,-1} \vb{e_z}) 
    \nonumber \\
    &+ \mathrm{Er} \frac{\pi}{4} \left(\frac{R}{r}-\frac{R^3}{r^3}\right) \left( \left( - \lambda  \Tilde{Y}_{2,2}+ \frac{1}{\sqrt{3}}  \Tilde{Y}_{2,0}\right)\vb{e}_x+ \left(-\frac{(\lambda+1)}{2}   \Tilde{Y}_{2,1}  \right) \vb{e}_z \right).
\end{align}
We introduced angular part of the real spherical harmonics as $Y_{1,l}=\sqrt{\frac{3}{4\pi}} \tilde{Y}_{1,l}$, where $l\in\{-1,0,1\}$ and $Y_{2,l}=\frac{1}{4}\sqrt{\frac{15}{\pi}} \tilde{Y}_{2,l}$, where $l\in\{-2,-1,0,1,2\}$.
We use the coefficients for the equilibrium dipolar and quadrupolar magnitudes, $\alpha$ and $q$, respectively, from numerical results in Ref.~\cite{stark_physics_2001}. To plot the structures containing defects (Fig. S6), we use the electrical field method described in \cite{stark_physics_2001}.

A second order solution can be obtained from Eq.~\ref{second-order-equation}, which can be viewed as a Poisson equation with terms in the form
$
    \nabla^2 \vb{n}^p_2(r,\theta,\phi)=\frac{\vb{\rho} (\theta,\phi)}{r^n}.
$
We can write the solution of such terms using spherical harmonics
$$
    \vb{n}^p_2(r,\theta,\phi)=-r^{2-n}\sum_{l=1}^{\infty}\sum_{m=-l}^{l} \frac{ \mathrm{Y}_{l}^{m}  }{(2-l-n)(l+3-n)}\bra{\vb{\rho}(\theta, \phi)}\ket{\mathrm{Y}_{l}^{m}(\theta, \phi)},
$$
where $\braket{f(\theta, \phi)}{g(\theta, \phi)}=\oint f(\theta, \phi) g(\theta, \phi) \sin\theta \dd{\theta} \dd{\phi}$. The result is a particular solution up to $\mathcal{O}(l=3)$ in spherical harmonics and uses Ericksen number as a parameter
\begin{align*} \delta^2_2 \vb{n}^{p} =&\; \Biggl[ q\,\mathrm{Er}\, \frac{(r-R)R^3}{336\,r^4} (1+7\lambda)\bigl(1+3\cos(2\theta)\bigr) \\ &\qquad +q\,\mathrm{Er}\, \frac{(r-R)R^3}{168\,r^4} (-31+3\lambda)\cos(2\phi)\sin^2\theta \\ &\qquad +\alpha\,\mathrm{Er}\, \frac{(r-R)R^2}{40\,r^3} (-13+15\lambda)\sin\theta\,\sin\phi \\ &\qquad +\mathrm{Er}^2\, \frac{\pi (r-R)R^2(-3r+2R)}{1344\,r^4} \bigl(-3+\lambda(-40+3\lambda)\bigr) \sin^2\theta\,\sin(2\phi) \Biggr]\vb{e}_x \\ &- \Biggl[ q\,\mathrm{Er}\, \frac{(r-R)R^3}{168\,r^4} (17+9\lambda)\sin(2\theta)\cos\phi \\ &\qquad +\mathrm{Er}^2\, \frac{\pi (r-R)R^2(-3r+2R)}{84\,r^4} (1+\lambda)\sin(2\theta)\sin\phi \Biggr]\vb{e}_z. \end{align*}
We add a homogeneous solution up to second order that will just try to balance these corrections out on the surface as $\delta^2 \vb{n}^p_2 (r=R)- \delta^2 \vb{n}^h_2(r=R) = 0$. This leads to the final expression $\delta^2 \vb{n}_2 = \delta^2 \vb{n}^h_2 +\delta^2 \vb{n}^p_2$, which we insert in the second order director approximation
\begin{align}
   \vb{n}\approx \vb{n_0}(1-\frac{1}{2} |\delta \vb{n_1}|^2)+\delta \vb{n_1} + \delta^2 \vb{n_2}. \label{second-order-director}
\end{align}
The cross-section of the final form of solutions is presented in Figure \ref{fig:director-structure}. To establish the behavior of the defect we follow the electric field representation technique introduced in \cite{stark_physics_2001}.

\noindent \subsection*{Propulsion of rotating spheres in nematic fluids}

To calculate the propulsion for the sphere we have to construct an appropriate stress tensor $\Sigma_{ij}$. The expression we use is the Ericksen-Leslie form of nematic stress tensor
\begin{equation}
    \Sigma_{ij}=\Sigma^\text{viscous}_{ij} + \Sigma^\text{Ericksen}_{ij}
    \label{eq:Sigma}
\end{equation}
Both stresses are defined as
\begin{align}
     \Sigma^\text{viscous}_{ij} &=\alpha_1 n_i n_j n_k A_{kl} n_l 
+ \alpha_2 n_i N_j 
+ \alpha_3 n_j N_i 
+ \alpha_5 n_i A_{jk} n_k 
+ \alpha_6 n_j A_{ik} n_k 
\label{s_viscous}
     \\
     \Sigma^\text{Ericksen}_{ij} &= - K \partial_i n_k\partial_j n_k + K \partial_k n_l \partial_k n_l \delta_{ij},\label{s_Er}
\end{align}
where we use the definitions for $A_{ij}=\frac{1}{2}(\partial_i v_j + \partial_j v_i)$ and $N_i = \dot{n}_i+v_j\partial_j n_i+\Omega_{ij}n_j =\frac{1}{\gamma_1} h_i - \frac{\gamma_2}{\gamma_1} \left( A_{ij} n_j - n_i (n_k A_{kj} n_j )\right)$ and $h_i = K (\partial_k \partial_k n_i -n_i n_j\nabla^2 n_j)$.

Using the above solution for the director field in combination with the homogeneous far field and the aligned dipole, we obtain the swimming velocity in the $x-$ and $y$-direction, respectively
\begin{align}
    U_x =\;& \frac{\alpha\,\mathrm{Er}\,K}{529200\,\gamma\mu R}
        \Bigl[8q\delta\bigl(-2260\alpha_1-2050\alpha_2+740\alpha_3+750\alpha_5
        -60\alpha_6+217\gamma+445\gamma\lambda\bigr) \nonumber\\
    &+ 315\Bigl(7\bigl(15\alpha_2+3\alpha_3-19\alpha_5+17\alpha_6
        -\pi\gamma(5+9\lambda)\bigr)-4\alpha_1\Bigr)\Bigr]
        + \mathcal{O}(\delta^4), \label{eq:Ux_general}\\[8pt]
    U_y =\;& \frac{\alpha\,\mathrm{Er}^2\,K}{21168000\,\gamma\mu R}
        \Bigl[(945\lambda-819)\bigl(17\alpha_2-165\alpha_3+73\alpha_5-137\alpha_6\bigr) \nonumber\\
    &+ 2\pi\Bigl(34880\alpha_2+80\bigl(187\alpha_3-342\alpha_5+459\alpha_6\bigr)
        +8317\gamma+240\lambda\bigl(140\alpha_2+125\alpha_3-346\alpha_5+365\alpha_6\bigr) \nonumber\\
    &\qquad\quad +32914\gamma\lambda-5595\gamma\lambda^2\Bigr)
        -64\alpha_1\bigl(270\lambda+5\pi(625\lambda+93)-234\bigr)\Bigr]
        + \mathcal{O}(\delta^4). \label{eq:Uy_general}
\end{align}
We can simplify these expressions by grouping terms into Ericksen number and 
the simplified stress tensor, we directly obtain the relations between Leslie viscosities $\alpha_i$ and alignment parameter $\lambda=-\frac{\gamma_2}{\gamma_1}$ and rotational viscosity $\gamma_1$
\begin{equation}
\begin{split}
\alpha_1&=\gamma_1\lambda^2,\qquad \alpha_2=\frac{1-\lambda}{2}\gamma_1 \qquad \alpha_3=-\frac{\lambda+1}{2}\gamma_1\\
\alpha_5&=\frac{\lambda-1}{2}\gamma_1\lambda, \qquad \alpha_6=\frac{\lambda+1}{2}\gamma_1\lambda.
\end{split}
\end{equation}
The final result for the velocity components is
\begin{align}
    U_x&= -\alpha \text{Er}\frac{ K (7 \pi  (9 \lambda +5)+\lambda  (11 \lambda -63)-42)}{1680 \mu  R}+\alpha q  \text{Er} \frac{ K (5 (139-383 \lambda ) \lambda -1178) }{66150 \mu  R} + \mathcal{O}(\delta^4), \\
    U_y &= -\alpha  \text{Er}^2 \frac{ K (9 (15 \lambda -13) (\lambda  (352 \lambda +217)-637)+2 \pi  (\lambda  (5 \lambda  (19544 \lambda -7545)-41834)-18277))}{21168000 \mu  R} + \mathcal{O}(\delta^4).
    \label{eq:propulsion_speed}
\end{align}
To compare with experiments we model configuration with the defect on top of the sphere, meaning that the dipolar vector $\vb{p}$ points along the $y$-axis. This corresponds to the choice of sign $\alpha < 0$ in our framework, which is reflected in the plot of our director solutions in Fig. \ref{fig:director-structure}. The magnitudes of the coefficients are obtained from \cite{stark_physics_2001}. We observe that to the first order in our expansion the swimming is mainly oriented in the $x$-direction and the velocity is proportional to the Ericksen number $U_x \propto \mathrm{Er}$. Correcting terms come in both $x$ and $y$ direction and create an angle at which the colloids start moving. In the $y$-direction the leading order is proportional to Ericksen squared $U_y \propto \mathrm{Er}^2$. In Figure S7, we show the swimming velocity for different flow-aligning parameters. Particularly, the propulsion component along the dipole $U_y$ has an opposite direction in flow-tumbling materials, compared to flow-aligning materials.

For the reciprocal motion we assume simplified model of a sinusoidal oscillation around the $z$-axis, in which the director structure adiabatically follows the rotation rate of the sphere. Upon averaging over a period of oscillation only $U_y$ remains and gains additional scaling factor of $1/2$.

\subsection*{Data availability}
All data needed to evaluate the conclusions in the paper are present in the paper and/or the Supplementary Materials.

\subsection*{Acknowledgments}
Work by M. Rajabi, Q.X.Z., Z.Z. and K.J.S. was supported by the U.S. Department of Energy (DOE), Office of Science, Basic Energy Sciences (BES) under Award DE-SC0022892.  S.A.C. acknowledges support through the Vagelos Institute for Energy Science and Technology Fellowship. 
Work by E.C., M.Ravnik, and Ž.K. was supported by the Slovenian Research and Innovation Agency (ARIS) under Grants P1-0099 and J1-50006, UL start-up programme under contract no. SN-ZRD/22-27/510-COMGATES, and the European Research Council under the Horizon 2020 Research and Innovation Program of the European Union (Program agreement 884928-LOGOS).

\subsection*{Authors contributions}
M. Rajabi helped to conceive the study, performed the experiments and analyzed the data. Q.X.Z. and Z.Z. helped with experiments. S.A.C. helped with experimental data analysis. E.C. and Z.K. developed the theory. G.P.A. helped with theory. M. Ravnik, Z.K., and K.J.S. conceived the research and led the project. All co-authors participated in scientific discussion and writing the manuscript.
\subsection*{Competing interests} 
The authors declare no competing interests.

\subsection*{Additional information}
{\bf Supplementary information} The submitted version contains supplementary material.

\newpage

% Figures, tables, equations and pages in the supplement are numbered S1, S2 etc.
\renewcommand{\thefigure}{S\arabic{figure}}
\renewcommand{\thetable}{S\arabic{table}}
\renewcommand{\theequation}{S\arabic{equation}}
\renewcommand{\thepage}{\arabic{page}}
\setcounter{figure}{0}
\setcounter{table}{0}
\setcounter{equation}{0}
\renewcommand{\thesection}{S\arabic{section}}
\setcounter{section}{0}
\setcounter{page}{1} % not 0 as \newpage already started a supplementary page
\pagenumbering{gobble} 
% References continue the numbering from the main text.
\begin{center}
    \section*{Supplementary information: Scallop Theorem for Swimming in Anisotropic Fluids}

    \author{
    	Mojtaba~Rajabi$^{\dagger}$, Enej Caf$^{\dagger}$, Qi Xing Zhang, Stephen A. Crane, Zhijian Zhou, Miha Ravnik, Gareth P. Alexander, Žiga Kos$^{\ast}$,
    	Kathleen~J. Stebe$^{\ast}$
       \\
    \small$^\ast$Corresponding authors. Email: ziga.kos@fmf.uni-lj.si, kstebe@seas.upenn.edu \\ 
    	\small$^\dagger$These authors contributed equally to this work.
        }
\end{center}

\setcounter{tocdepth}{3}
\renewcommand{\contentsname}{Table of Contents}
\tableofcontents

\renewcommand{\thepage}{\arabic{page}}
\setcounter{page}{1}
\newpage

\section{Supplemental figures}

\begin{figure}[h!]
	\centering
	\includegraphics[width=0.85\textwidth]{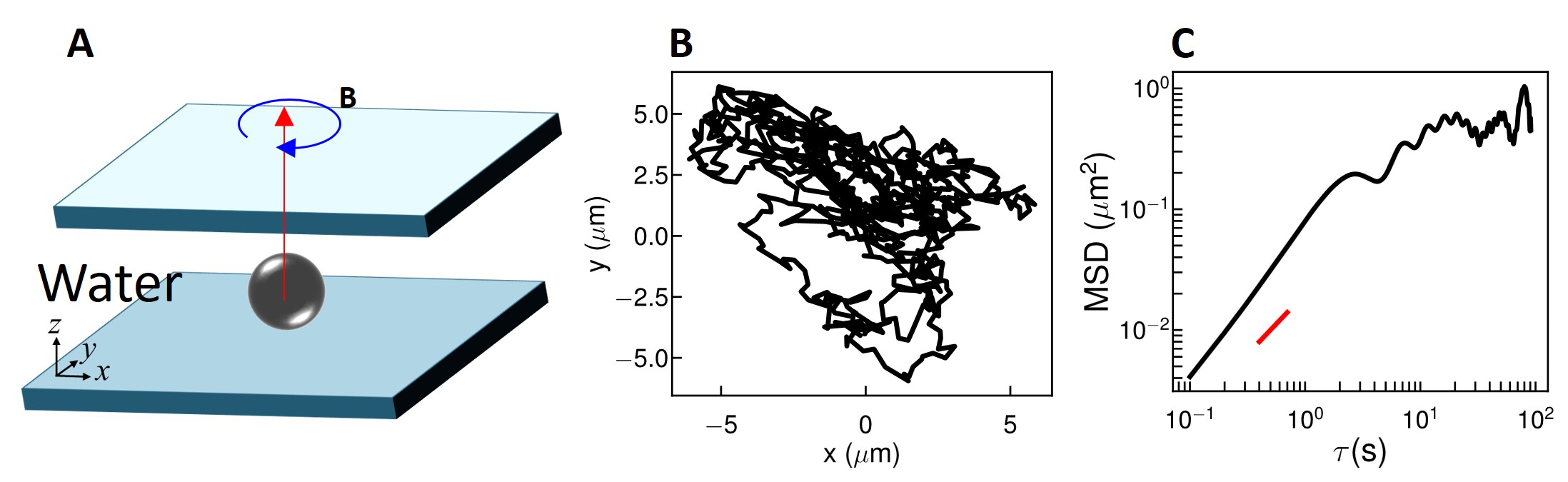} 
	\caption{\textbf{~Diffusive motion of a rotating a sphere in an isotropic liquid. (A)} Schematic of a rotating ferromagnetic sphere in a cell filled with water.  (\textbf{B}) Trajectory of a sphere rotating with $T=10$ s in water over 100 s. (\textbf{C}) MSD of the rotating sphere as a function of lag time $\tau$.}
	\label{Sup water rotation} 
\end{figure}
\clearpage
\newpage
\vspace*{-2cm}
\begin{figure}[t!] 
	\centering
	\includegraphics[width=0.85\textwidth]{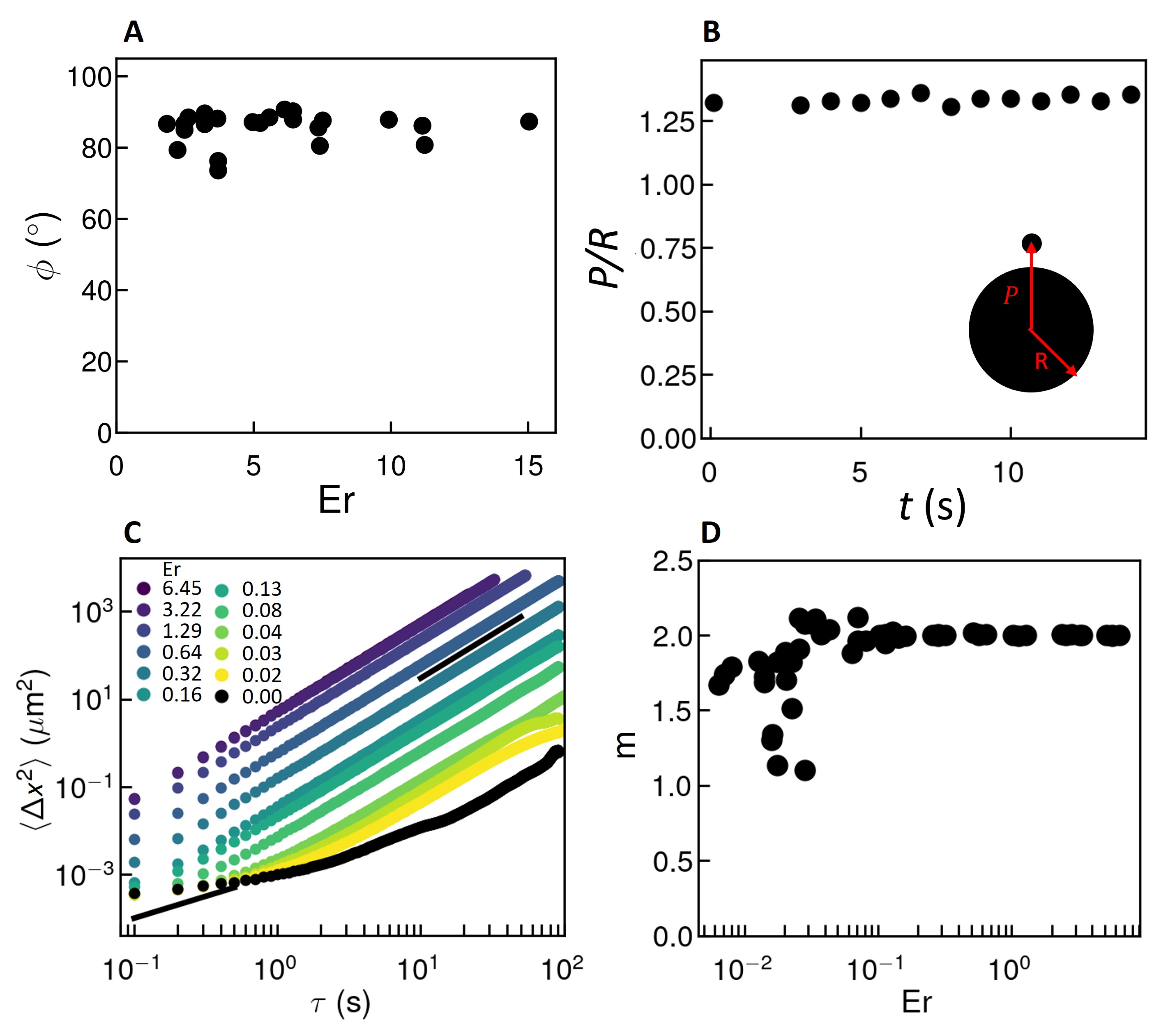}
	\caption{\textbf{~Rotating sphere in 5CB.}(\textbf{A}) The angle $\phi$ between the sphere trajectory and $y$-axis. The rotating spheres translate predominantly perpendicular to the nematic director orientation. (\textbf{B}) The plot shows the ratio of the distance between the point defect and the center of particle P to the sphere radius R as a function of time. It indicates that the defect remains in the original plane, confirming the absence of any out-of-plane rotation of the sphere. (\textbf{C}) MSD vs $\tau$. (\textbf{D}) The exponents of MSD versus Er.}  
	\label{Sup 5CB rotation} 
\end{figure}
\clearpage
\newpage
\vspace*{-2cm}
\begin{figure} [t!]
	\centering
	\includegraphics[width=0.85\textwidth]{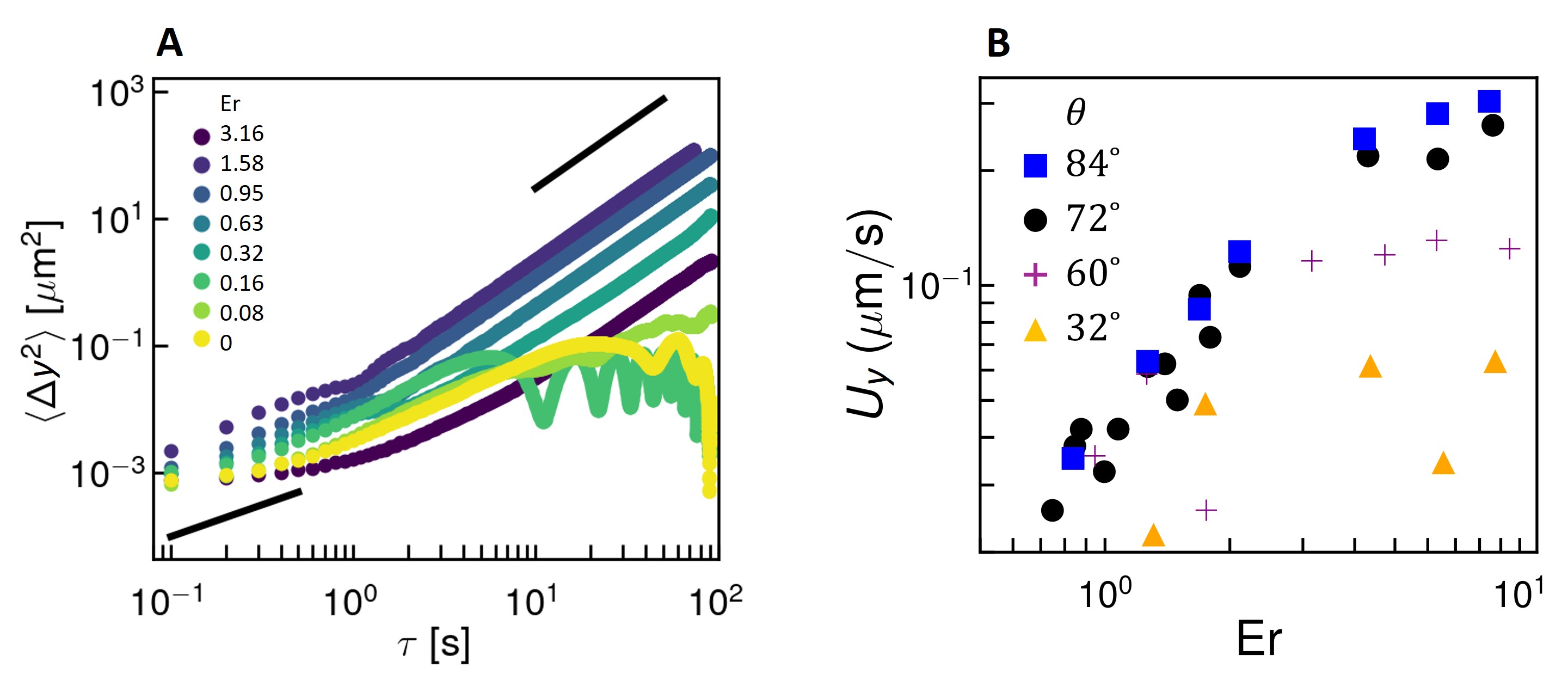}
	\caption{\textbf{~Reciprocal forcing of sphere in 5CB.}(\textbf{A}) MSD vs $\tau$. (\textbf{B}) Speed along the director as a function of Er for forcing with different angles.} 
	\label{Sup 5CB Oscillation} 
\end{figure}
\clearpage
\newpage
\vspace*{-2cm}
\begin{figure} [t!]
	\centering
	\includegraphics[width=0.8\textwidth]{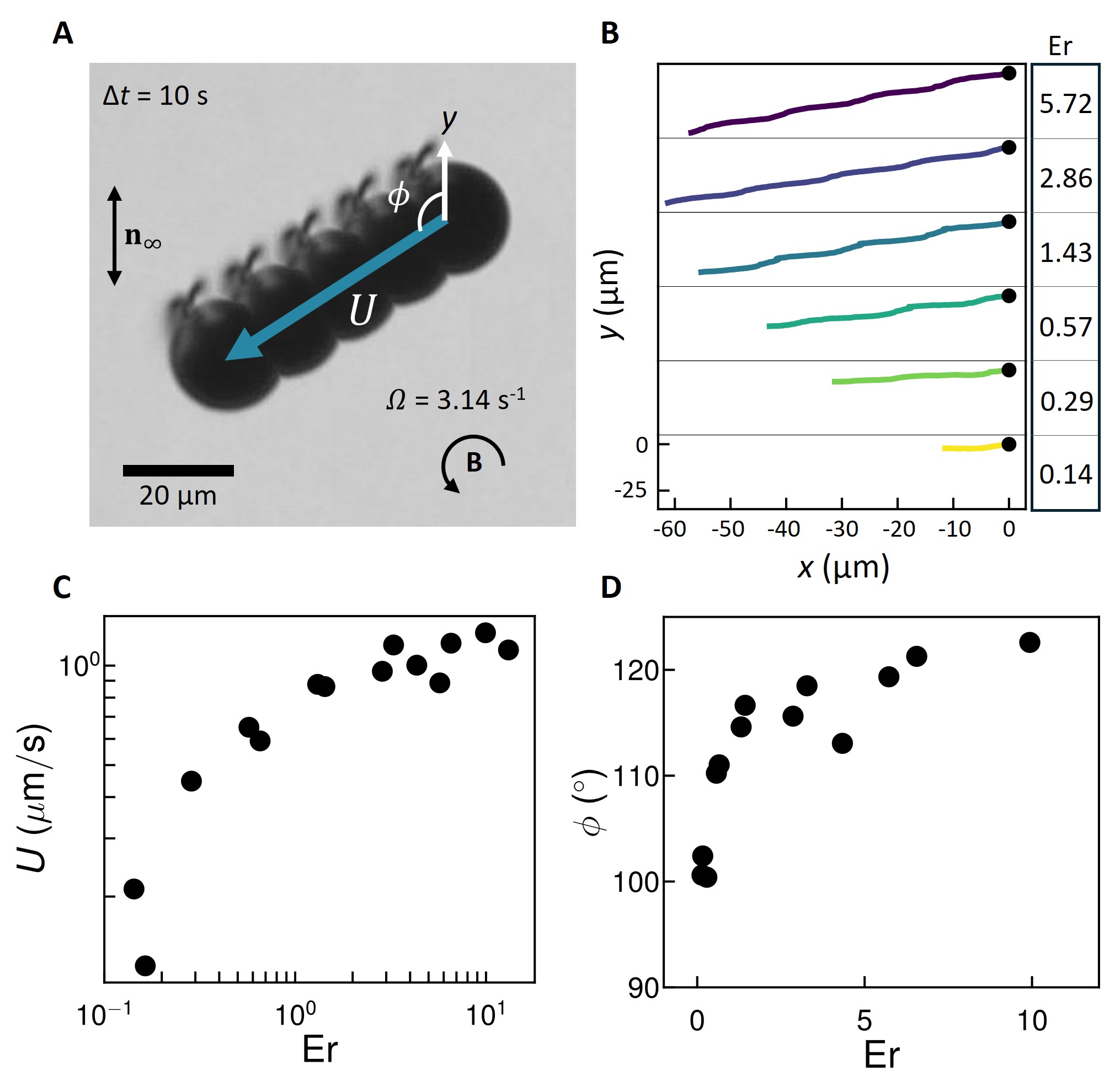}
	\caption{\textbf{~Swimming of a rotating sphere  in a mixture of 7CB and 8CB.}(\textbf{A}) Overlay of microscopy textures of a rotating sphere in a mixture of 7CB and 8CB. The arrow represents the spheres' swimming trajectory. $\Delta t$ is the time interval between two consecutive overlaid images. (\textbf{B}) Trajectories of the rotating sphere at different Er over a duration of $t=20$~s. (\textbf{C}) Speed of rotating spheres in a mixture of 7CB and 8CB as a function of Er. (\textbf{D}) The angle $\phi$ between the trajectory of sphere and $y$-axis as a function of Er.}. 
	\label{Sup 7CB Rotation} 
\end{figure}
\clearpage
\newpage
\vspace*{-2cm}
\begin{figure} [t!]
	\centering
	\includegraphics[width=0.85\textwidth]{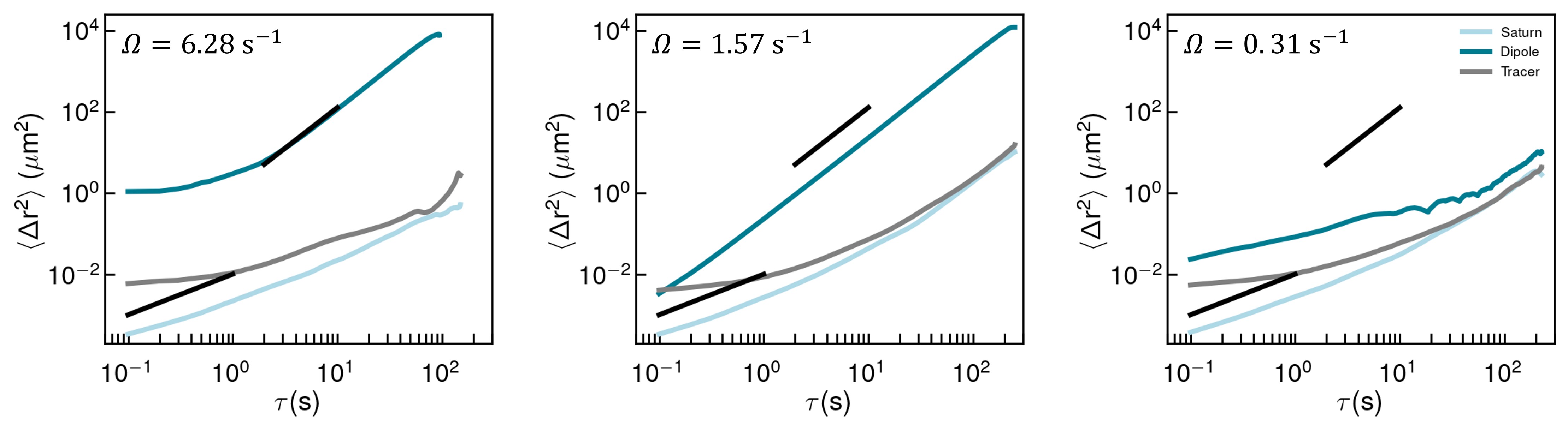}
	\caption{\textbf{~Rotating sphere with Saturn ring disclination in 5CB moves with attentuated velocity comparable to drift.} Comparison of MSD vs $\tau$ for spheres with dipole and Saturn ring configurations at different rotational speeds. The displacement of the sphere with the Saturn ring is insignificant and comparable to the drift of a tracer particle.} 
	\label{sup saturn ring} 
\end{figure}
\clearpage
\newpage
\vspace*{-2cm}
\begin{figure} [t!]
	\centering
	\includegraphics[width=0.66\textwidth]{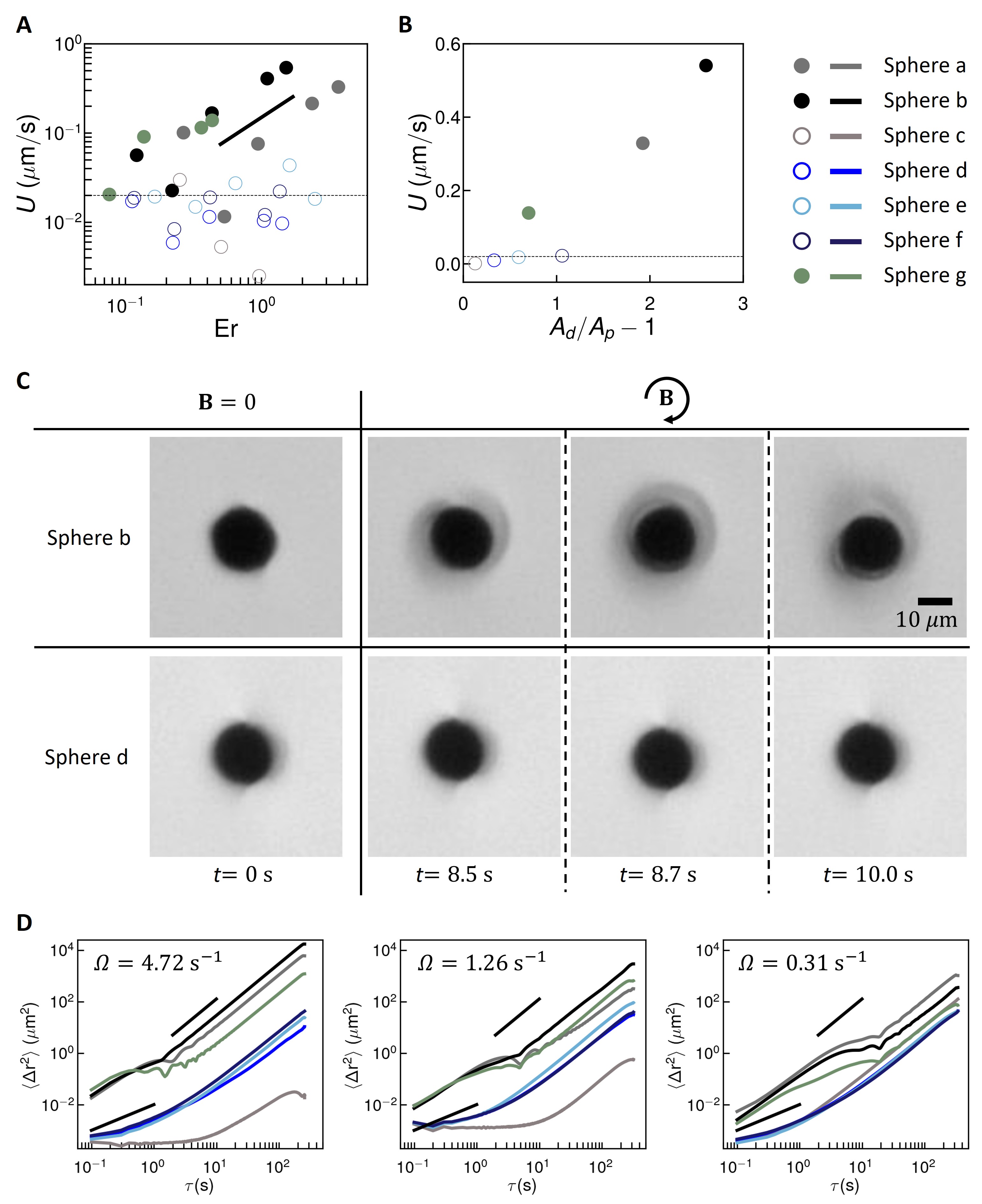}
	\caption{\textbf{~Rotating tangentially anchored sphere in 5CB.} \blue{(\textbf{A}) Translational speed as a function of $\mathrm{Er}$ for different spheres. Several spheres propel under rotation, whereas the remaining spheres exhibit predominantly diffusive motion, with speeds comparable to the background drift indicated by the dotted line. (\textbf{B}) Translational speed as a function of the area of the distorted nematic region surrounding each sphere. Spheres that strongly distort the surrounding nematic generally propel faster. (\textbf{C}) Snapshots of a motile sphere (top row) and a stationary sphere (bottom row) at equlibrium and under rotation with $\mathit\Omega=4.72$~s$^ {- 1}$ . The motile tangentially anchored sphere strongly distorts the surrounding nematic and shows complex defect dynamics, whereas the stationary sphere produces only weak distortions. (\textbf{D}) MSD vs $\tau$, for multiple spheres at different rotational speeds. The weakly distorting spheres are predominantly diffusive with speed comparable to the background drift, whereas strongly distorting spheres undergo translational motion. }} 
	\label{sup boojum} 
\end{figure}

\clearpage
\newpage
\vspace*{-2cm}
\begin{figure}[t!]
	\centering
	\includegraphics[width=0.85\textwidth]{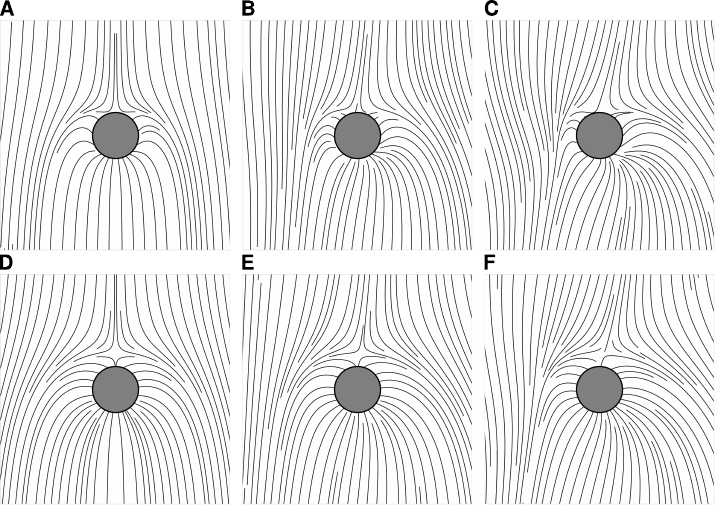} 
	\caption{\textbf{~Analytically obtained director profiles for various $\mathrm{Er}$.}
    We show cross section plot of $\vb{n}\approx \vb{n_0}(1-\frac{1}{2} |\delta \vb{n_1}|^2)+\delta \vb{n_1} + \delta^2 \vb{n_2}$ up to \blue{second} order in $\delta$. We used parameters $|\alpha|= 3.08$, $|q|=1.5$ and $\lambda=1.05$ based on \cite{stark_physics_2001} for (\textbf{A}) $\text{Er}=0$, (\textbf{B}) $\text{Er}=1$ and (\textbf{C}) $\text{Er}=2$. In addition to the multipolar framework we also plot the electric field inspired construction \cite{stark_physics_2001} for the corresponding second order director field around the colloid, which shows the location of the defect for (\textbf{D}) $\text{Er}=0$, (\textbf{E}) $\text{Er}=1$ and (\textbf{F}) $\text{Er}=2$.}
	\label{fig:director-structure} 
\end{figure}
\clearpage

\clearpage
\newpage
\vspace*{-2cm}
\begin{figure}[t!]
	\centering
	\includegraphics[width=0.85\linewidth]{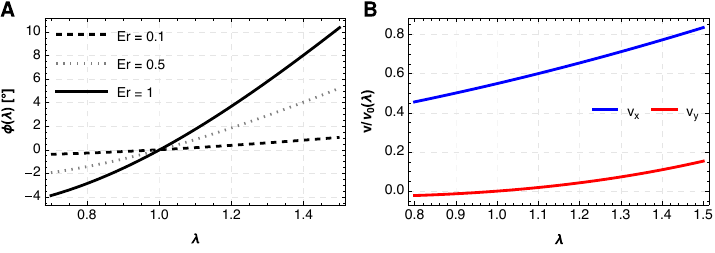}
    \caption{\textbf{~Dependence of swimming angle and velocity on material alignment parameter.}~(\textbf{A}) Angle of swimming depending on alignment parameter at three different Ericksen numbers. We can observe that the sign changes at $\lambda \approx 1.0018$, which also distinguishes between the flow alignment and tumbling. (\textbf{B}) Components of swimming velocity depending on the alignment parameter. They are all scaled with $v_0 = \frac{K}{R \mu}$.}
    \label{fig:swimming-direction}
\end{figure}

\clearpage
\newpage

\section{List of Supplementary movies}
\paragraph{Caption for Movie S1.}
\textbf{Diffusive motion of a rotating ferromagnetic spherical colloid in an isotropic liquid.}
\paragraph{Caption for Movie S2.}
\textbf{Swimming of a rotating spherical colloid in nematic liquid crystal 5CB.} The sphere translates perpendicular to the nematic far field director. 
\paragraph{Caption for Movie S3.}
\textbf{Swimming of spherical colloids with a reciprocal driving force in nematic liquid crystals 5CB.} The sphere translates along the far field director while the point defect leads the way. 
\paragraph{Caption for Movie S4.}
\textbf{Swimming of a rotating spherical colloid in a mixture of 7CB and 8CB.} The sphere translates with an angle $>90^\circ$ with respect to the far field director. 
\paragraph{Caption for Movie S5.} \textbf {Swimming of spherical colloids with a reciprocal driving force in a mixture of 7CB and 8CB.} The sphere translates along the far field director while the sphere leads the way and the point defect follows.
\paragraph{Caption for Movie S6.}
\textbf{Rotation of a spherical colloid with Saturn ring defect structure in a nematic liquid crystal 5CB.} Quadrupolar symmetry of the nematic structure around the large sphere prevents effective swimming due to non-broken symmetries, while a neighboring small sphere with dipolar defect swims effectively.

\section{ Propulsion of swimmers in complex fluids}

Here we provide a general framework and show how temporal and spatial symmetries of the swimmer's body and the structure of the surrounding fluid enable or prohibit propulsion of microswimmers. This is done by combining the Lorentz reciprocal theorem for complex fluids adopted from \cite{neville2024controlling} and temporal/spatial symmetry conditions, where we follow the main ideas from \cite{Lauga2009}. In Section \ref{SI:arbitrary-fluids}, we focus on arbitrary complex fluids described by an additional contribution to the stress tensor. In Section \ref{SI:nematic-fluids} we focus specifically on nematic fluids and show how propulsion velocity scales with Ericksen number for a swimmer with an arbitrary shape.

\subsection{Breaking of temporal and spatial symmetries} \label{SI:arbitrary-fluids}
We start by defining an arbitrary swimmer, whose exterior defines the domain $D=\{(x,y,z)\in  \mathbb{R}^3 \}$, representing the bulk around the swimmer. The swimmer of interest positioned in the complex fluid is characterised with the following boundary condition on its surface $\mathbf{v}\big|_{\partial D}
  = \mathbf{U} + \boldsymbol{\Omega}\times\mathbf{x} +
  \mathbf{u}_S,$
where $\mathbf{u}_S$ is a velocity on the swimmer's surface $\partial D$ with normal $\bm{\nu}$, and $\vb{U}$ and $\vb{\Omega}$ denote the velocity and rotation rate, while $\vb{x}$ denotes the position in the bulk. We proceed with the Lorentz reciprocal theorem \cite{Lauga2009}, a common tool in hydrodynamics of microswimmers used to derive desired solutions through convenient use of known (or easier to extract) solutions of standard hydrodynamic problems for an auxiliary passive swimmer $\mathbf{v}'\big|_{\partial D}
  = \mathbf{U}' + \boldsymbol{\Omega}'\times\mathbf{x}$.

To implement the reciprocal theorem, we define the Stokes equation for the structured fluid with pressure $p$ and flow $\vb{v}$ with additional generic anisotropic stress tensor $\vb{\Sigma}$, and a Stokes equation for the auxiliary problem in isotropic fluid with pressure $p'$ and $\vb{v'}$ as
\begin{align}
    \mu \nabla^2 \vb{v} - \nabla p + \nabla \cdot \vb{\Sigma}&=0, \\
    \mu \nabla^2 \vb{v}' - \nabla p' &=0.
\end{align}
The anisotropic and isotropic fluids have the following corresponding stresses $\sigma_{ij} = \mu (\partial_i v_j + \partial_j v_i) - p \delta_{ij} + \Sigma_{ij}$ and $\sigma'_{ij} = \mu (\partial_i v'_j + \partial_j v'_i) - p' \delta_{ij}$. 
Motivated by~\cite{neville2024controlling}, we proceed with the Lorentz reciprocal theorem, and write
\begin{equation}
    \int_D \partial_i (v'_j \sigma_{ij}) - \partial_i (v_j \sigma'_{ij}) \dd{V} = \int_{\partial D} {u_S}_i  \sigma'_{ij} \nu_j \dd{S} -\int_{D} \Sigma_{ij} \partial_i v'_j \dd{V}, \label{Reciprocal-integral-form}
\end{equation}
Using the divergence theorem, we can evaluate the surface integrals on the left-hand-side of the Equation \ref{Reciprocal-integral-form}. We find relations between rotational and translational velocities with forces and torques in both systems
\begin{equation}
   \Omega'_i T_i + U'_i F_i - \Omega_i T'_i - U_i F'_i = \int_{\partial D} {u_S}_i  \sigma'_{ij} \nu_j \dd{S} -  \int_D \Sigma_{ij} \partial_i v'_j \dd{V}.
\end{equation}
We can now focus on torque and force free swimmers, extracting the propulsion velocity
\begin{equation}
   U_i F'_i+\Omega_i T'_i = - \int_{\partial D} {u_S}_i  \sigma'_{ij} \nu_j \dd{S}+ \int_D \Sigma_{ij} \partial_i v'_j \dd{V}.
   \label{eq:Lorentz-1}
\end{equation}
We see that the main novelty to the traditional result in \cite{stone1996propulsion} is evaluating the contribution arising from the additional stress tensor. We can see that the way to generate the net propulsion is to make the integral terms non-zero through shape, stroke and fluid structure.

To resolve the integrals in Eq.~\ref{eq:Lorentz-1} we separate different components of the contracted forms. For the propulsion component along the $k$-th direction, we choose the auxiliary flow field $\vb{v}'^{(k)}$ corresponding to a rigid translation in that direction. The reciprocal theorem then separates naturally into an traditional isotropic contribution and an additional contribution $\Sigma_{ij}$ arising from complex nature of the fluid.

Similarly, the corresponding propulsion velocity can be decomposed as
\begin{equation}
    U_k(t)
    =
    U_k^{\mathrm{Newtonian}}(t)
    +
    U_k^{\mathrm{complex}}(t),
\end{equation}
where
$
U_k^{\mathrm{Newtonian}}
=
-
M_k
\int_{\partial D}
{u_S}_i
\sigma'^{(k)}_{ij}\nu_j
\dd S
$
is the usual Newtonian isotropic contribution and
$
U_k^{\mathrm{complex}}
=
M_k
\int_{D}
\Sigma_{ij}
\partial_i v_j'^{(k)}
\dd V
$
is the contribution arising from the additional stress tensor. Here $M_k$ denotes the mobility coefficient associated with the Newtonian and complex non-Newtonian responses. Time dependence is contained in the velocity $\vb{u}_S$ prescribed on the surface $\partial D$ and complex stresses in the bulk $\Sigma_{ij}$, but could also show in a time dependent mobility for a highly deformable swimmer.

We take the stroke to be a periodic series of actions that the swimmer performs, which results in defining a natural period of motion $t=T$. Under this notation the net displacement may therefore be decomposed as
\begin{equation}
    d_k
    =
    d_k^{\mathrm{Newtonian}}
    +
    d_k^{\mathrm{complex}},
\end{equation}
with
$
d_k^{\mathrm{Newtonian}}
=
\int_0^T U_k^{\mathrm{Newtonian}}\,\dd t
$
and
$d_k^{\mathrm{complex}}
    =\int_0^T U_k^{\mathrm{complex}}\,\dd t=
    \int_0^T
    M_k
    \int_{D}
    \Sigma_{ij}
    \partial_i v_j'^{(k)}
    \dd V
    \dd t .$
The central question then becomes determining under which temporal and spatial symmetries the functional $M_k
\int_{D}
\Sigma_{ij}
\partial_i v_j'^{(k)}
\dd V$ vanishes identically, vanishes only after integration over a full $T$-cycle, or survives to yield a finite contribution to propulsion. The most important part in exploring the swimming criteria is to look at the temporal symmetry, however spatial symmetries can indeed play a role and prevent too spatially symmetric swimmers and their responses in the complex fluids to move.

Firstly, in order to investigate vanishing of $d_k^{\mathrm{complex}}$ over a full cycle, we proceed to define time reversal operator $\mathcal{T}$, which for a $T$-periodic field $\vb{f}(\vb{x},t)$ can be defined as $\mathcal{T}\vb{f}(\vb{x},t)=\vb{f}(\vb{x},-t) $. We can find that if a $T$-periodic field is odd under time reversal $\mathcal{T}\vb{f}(\vb{x},t)=\vb{f}(\vb{x},-t)=-\vb{f}(\vb{x}, t) $ then the integral of it vanishes, which can be seen by the change of variables $t $ to $-t$ together with periodicity. In our case we investigate, whether the $T$-periodic $U^{\mathrm{complex}}_k$ is odd under time reversal or not, which will depend based on the choice of the complex stress tensor $\Sigma_{ij}$.

Secondly, to investigate the spatial symmetry requirements for propulsion, we have to consider the bulk domain $D$. Note that we defined it as a subset of $\mathbb{R}^3$, but the same formalism is also applicable for the two dimensional hydrodynamics ($D \subset \mathbb{R}^2$). We define integrand $\mathcal{I} : D \to \mathbb{R}$ to be an integrable scalar, where we take $ \mathcal{I} =\Sigma_{ij} \partial_i v'^{(k)}_j$, belonging to the displacement into the $k$-th direction in space. We now suppose that there exists a transformation $\mathcal{S} : \mathbb{R}^n \to \mathbb{R}^n$ under which the domain is invariant $\mathcal{S}(D)=D$, it preserves volume $|\det(\nabla \mathcal{S})|=1$ and the integrand is odd under it $\mathcal{I}(\mathcal{S} \vb{x})=- \mathcal{I}(\vb{x})$ for every $\vb{x}\in D$. If such transformation exists then $I=\int_D \mathcal{I}(\vb{x}) \dd{V}=0$. The proof can be done through a change of variable, and use of invariance of the domain with oddness of the integrand under transformation $\mathcal{S}$. For time dependent swimmers this has to be checked at every instant.

To find the transformation $\mathcal{S}$, we can look for rotations, inversions and reflections. Any such transformation implies the vanishing of the corresponding spatial integral. Reflections constitute a particularly useful class of transformations since they naturally generate sign changes in scalar quantities. For a given propulsion component, one may choose the reflection most convenient for analysing the symmetry properties of the integrand based on the auxiliary flow field. We note that especially in complex fluids, the reflection generating the symmetry breaking might not be in the direction of swimming, as we demonstrate in the paper for the rotating sphere with the dipolar structure.

One of the main consequences is also that the symmetry of the domain and the symmetry of the integrand under $\mathcal{S}$ are connected requirements. The first condition guarantees that every point in the bulk $\vb{x}\in D$ is paired with its transformed counterpart $\mathcal{S}\vb{x}\in D$, while the second guarantees cancellation between the two contributions. Neither condition alone is sufficient. For example, an odd function integrated over a domain that is not invariant under the corresponding transformation does not need to vanish. This is why for structured complex fluids looking at the oddness of the integrand through spatial symmetries of the fluid is especially convenient for swimmers that are highly symmetric under reflections such as sphere, ellipsoid, cylinder, torus etc. In the spatial part, we will only consider these, as they actually bring new physical aspect in terms of propulsion. Other non-reflection-symmetric swimmers already have their prepositions to break required spatial symmetries in the Newtonian fluids or viscoelastic fluids.

For our swimmers we now conclude that both temporal and spatial operators are involutions $\mathcal{T}^2=\mathcal{S}^2=\mathbb{I}$ that commute, meaning that they generate $\mathbb{Z}_2 \times \mathbb{Z}_2$ group. We can now decompose every field into irreducible character pairs $(\pm, \pm)$ corresponding to the symmetry under $\mathcal{T}$ and $\mathcal{S}$ and study them separately, greatly simplifying the analysis going further. The criterion is: whether the integrand is odd under time reversal and whether the integrand is odd under some reflection, which obeys the symmetry of the domain (swimmer's shape). We can now look at one of the examples of structured fluids beyond Newtonian fluids and try to draw general conclusions.

\subsection{ Swimming in nematic fluids} \label{SI:nematic-fluids}
We establish general principles governing the propulsion velocity of swimmers with arbitrary shapes in the regime of low Ericksen number within the Ericksen-Leslie framework for nematic fluids.

\subsubsection{ Temporal symmetry}

First, we consider the director fields for which the zero-th order quasi-static solution
explicitly depends on the instantaneous shape of the swimmer. To describe the higher-order director field distortions around such swimmers, we decompose it as
\begin{equation}
    \vb{n} = \frac{\vb{n}_0(\vb{r},t)+\mathrm{Er}\vb{n}_1+...}{|\vb{n}_0(\vb{r},t)+\mathrm{Er}\vb{n}_1+...|}=\vb{n_0}+\mathrm{Er}\left(\vb{n_1} - \vb{n_0} (\vb{n_0}\cdot \vb{n_1})\right)+\mathcal{O}\left(\mathrm{Er}^2\right). \label{general-linear-expansion-n}
\end{equation}
where $\vb{n_0}$ represents the quasi-static director distortion determined by the instantaneous swimmer shape, and $\vb{n_1}$ denotes the first-order correction arising from the coupling to the generated flow field. Note that here we only consider $\mathrm{Er}$ as a small parameter, while viscous anisotropy can be arbitrary and multipolar expansion is not necessary.

To describe the dynamics of the system, we use the linear expansion (Eq. \ref{general-linear-expansion-n}) in the Eq.~\ref{director-Eq}. This yields a system of equations
{\small
\begin{align}
    &\frac{K}{\gamma_1} \left( \nabla^2 \vb{n_0}- \vb{n_0} (\vb{n_0}\cdot \nabla^2 \vb{n_0}) \right)=0 +\mathcal{O}\left(\mathrm{Er}\right), \\
     &\frac{K}{\gamma_1} \left( \nabla^2 \vb{n_1}-\nabla^2 \vb{n_0} (\vb{n_0}\cdot \vb{n_1}) -\vb{n_0} (\vb{n_0}\cdot \nabla^2 \vb{n_1}-\vb{n_0} (\vb{n_1}\cdot \nabla^2 \vb{n_0})-\vb{n_1} (\vb{n_0}\cdot \nabla^2 \vb{n_1} \right)=\nonumber \\  & \qquad =
     \partial_t \vb{n_0}
     + \vb{v} \cdot \nabla \vb{n_0}  +\vb{\Omega } \vb{n_0}- \lambda (\vb{A}\vb{n_0}-\vb{n_0}(\vb{n_0}\cdot \vb{A} \vb{n_0})) +\mathcal{O}\left(\mathrm{Er}^2\right).
     \label{first-order-n1}
\end{align}
}
Solving these equations necessarily depends on the specific swimmer's geometry, and we illustrate the procedure explicitly for spherical swimmers. In general, we can still observe that the quasi-static part $\vb{n_0}$ is determined solely by the swimmer’s shape, whereas the first-order correction $\vb{n_1}$ also depends on the applied forcing.

Having established the framework for computing the director field, we now proceed to the central question -- how such distortions translate into swimming motion. Following~\cite{Lauga2009}, we will consider propulsion due to contributions from the viscous stress tensor Eq. \ref{s_viscous} and the Ericksen stress tensor Eq. \ref{s_Er}. However, due to simplicity, we will treat them separately.

Let us start with the propulsion due to the viscous stress tensor Eq.~\ref{s_viscous}. By rescaling the time by a factor of $k$, $t\rightarrow t/k$, flow field solutions and the director corotational time derivative scale by the same factor
\begin{equation}
    \vb{v}\rightarrow k\vb{v}, \quad \vb{v'}\rightarrow k\vb{v'}, \quad\vb{U'}\rightarrow k\vb{U'}. 
\end{equation}
Using this we can show in the lowest order also the same holds for the corotational derivative. By inserting simple definitions we end up with
\begin{equation*}
    \vb{N}_0=\partial_t \vb{n}_0 + \vb{v} \cdot \nabla \vb{n}_0  +\vb{\Omega} \vb{n}_0 \rightarrow k \partial_t \vb{n}_0 + k\vb{v} \cdot \nabla \vb{n}_0  + k \vb{\Omega} \vb{n}_0 = k\vb{N}_0
\end{equation*}
The same can now be used to transform the stress tensor depending on $\vb{n_0}$, $\vb{N_0}$ and $\vb{v}$ as $\Sigma_{ij} \rightarrow k\Sigma_{ij}$ in the lowest order of the director expansion. 
For simplicity, we redefine the force dependence on the swimming velocity as $\vb{F'}=\xi \vb{V'}$, where for a spherical swimmer $\xi=-6\pi \mu R$. The same scaling with $k$ can be obtained for swimmers in the auxiliary problem, where the applied force is additionally proportional to the rotation rate. The above arguments lead to scaling of the propulsion velocity as
\begin{align}
    \vb{U} \cdot k\vb{V'} = \frac{1}{\xi} \int_D k\Sigma_{ij} \partial_i k v'_j \dd{V} \Rightarrow \vb{U} \rightarrow k \vb{U}.
\end{align}
This demonstrates that an instantaneous reversal of the forcing ($k=-1$) in the lowest order does not modify the structure of the flow solutions, but only the sign of the flow. This corresponds with the work by Purcell for the Stokes equations \cite{purcell2014life}, which state that the dynamic solutions, which are linearly scalable in time, cannot generate propulsion for reciprocal swimming strokes. This is true even if parts of the swimming stroke are performed at different rates. However, non-reciprocal swimming strokes can still generate motion, which will, provided that the reflection symmetry is broken, scale linearly in Ericksen number
\begin{equation}
    U\propto \mathrm{Er},
\end{equation}
however they are allowed to swim purely on the grounds of the time reversal symmetry also in the Newtonian fluids.  

If we include the director expansion to the first order in $\mathrm{Er}$, the viscous stress tensor (Equation~\ref{s_viscous}) can no longer be linearly rescaled and propulsion can occur even for reciprocal swimming strokes. The magnitude of such propulsion velocity is proportional to $\mathrm{Er} \, \vb{n}_1$  and the generated flow field (of order $\mathrm{Er}$), which leads to quadratic scaling in Ericksen number
\begin{equation}
    U \propto \mathrm{Er}^2.
\end{equation}
For completeness, we must also consider the contribution to the propulsion velocity due to the Ericksen stress (Equation~\ref{s_Er}). 
For $\vb{n}_0$, the director field is in equilibrium, in which case the elastic stresses cannot generate propulsion. The first order correction term $\vb{n}_1$ satisfies Eq.~\ref{first-order-n1} and scales as $\mathrm{Er} \,\vb{n_1}\rightarrow k \mathrm{Er} \, \vb{n_1}$, which means that propulsion due to the elastic stresses generated by the distorted director field can only occur as a second order effect in $\mathrm{Er}$.

\subsubsection{Spatial symmetry}
We established the leading order from time symmetry, however, the spatial symmetries can restrict this further. We note that any symmetric configuration of swimmer which will have the odd-behavior in the integrand under reflection will lead to no propulsion. For example a spherical swimmer in an isotropic fluid is not able to propel. This is well established and is also demonstrated in this study, where externally actuated spheres fail to generate net propulsion in a Newtonian fluid. We can then see that the symmetry arguments are also powerful in structured fluids, where we have to additionally look for the spatial symmetries of the director. The symmetric director field, for example a pure quadrupolar structure ($q\ne 0$, $\alpha=0$), will restrict propulsion and the swimmer will not move. This argument is  similar to the one in isotropic fluids, except that it additionally depends on the symmetries of the bulk, rather than just the surface of the swimmer. On the other hand for the dipole ($\alpha \ne 0$), which breaks the reflection symmetry, the situation is different. The object is allowed to swim nonreciprocally at linear order, with the Newtonian part being suppressed only because of the swimmer's symmetry, and will swim quadratically if the stroke is reciprocal. For most swimmers the easiest way to swim in complex fluid is to use a complex nonreciprocal swimming stroke  \cite{yao2022topological, zhang_sweeping_2026}, which is scaled linearly with Er.

We can conclude that the reported scaling of $U$ is universal for swimmers with arbitrary shapes in the regime of low Ericksen numbers in anisotropic fluids. The swimmers need to break the spatial and temporal symmetries that are summarized in Fig. \ref{figure 1}, which comes directly from the combination of the underlying swimmer, director and flow symmetries within the integrals in the Lorentz reciprocal theorem.

\end{document}